\documentclass{ieeeaccess}
\usepackage{cite}
\usepackage{amsmath,amssymb,amsfonts}
\usepackage{algorithmic}
\usepackage{graphicx}
\usepackage[font={sf,scriptsize},           % new
            labelfont={bf,color=accessblue},% new
            caption=false]                  % new
           {subfig} 
\usepackage{textcomp}

\usepackage[table]{xcolor}
\usepackage{multirow, booktabs}
\graphicspath{ {./Images/} }
\usepackage[utf8]{inputenc}
\usepackage{dirtytalk}
\usepackage{comment}
\usepackage{tabularx} % in the preamble
\usepackage{scalerel}
\usepackage{tikz}
\usetikzlibrary{svg.path}

\definecolor{orcidlogocol}{HTML}{A6CE39}
\tikzset{
  orcidlogo/.pic={
    \fill[orcidlogocol] svg{M256,128c0,70.7-57.3,128-128,128C57.3,256,0,198.7,0,128C0,57.3,57.3,0,128,0C198.7,0,256,57.3,256,128z};
    \fill[white] svg{M86.3,186.2H70.9V79.1h15.4v48.4V186.2z}
                 svg{M108.9,79.1h41.6c39.6,0,57,28.3,57,53.6c0,27.5-21.5,53.6-56.8,53.6h-41.8V79.1z M124.3,172.4h24.5c34.9,0,42.9-26.5,42.9-39.7c0-21.5-13.7-39.7-43.7-39.7h-23.7V172.4z}
                 svg{M88.7,56.8c0,5.5-4.5,10.1-10.1,10.1c-5.6,0-10.1-4.6-10.1-10.1c0-5.6,4.5-10.1,10.1-10.1C84.2,46.7,88.7,51.3,88.7,56.8z};
  }
}

\newcommand\orcid[1]{\href{https://orcid.org/#1}{\mbox{\scalerel*{
\begin{tikzpicture}[yscale=-1,transform shape]
\pic{orcidlogo};
\end{tikzpicture}
}{|}}}}

\usepackage{hyperref} 
\NewSpotColorSpace{PANTONE}
  \AddSpotColor{PANTONE} {PANTONE3015C} {PANTONE\SpotSpace 3015\SpotSpace C} {1 0.3 0 0.2}
\SetPageColorSpace{PANTONE}
\usepackage[framemethod=tikz]{mdframed}

\def\BibTeX{{\rm B\kern-.05em{\sc i\kern-.025em b}\kern-.08em
    T\kern-.1667em\lower.7ex\hbox{E}\kern-.125emX}}

\begin{document}
\mdfsetup{%
    linecolor=gray!50!white,
    % outerlinewidth=2pt,
    roundcorner=5pt,
    % innertopmargin=4pt,
    % innerbottommargin=4pt,
    % innerrightmargin=4pt,
    % innerleftmargin=4pt,
    %     leftmargin = 4pt,
    %     rightmargin = 4pt
    backgroundcolor=gray!10!white}

\history{This work has been submitted to the IEEE for possible publication. Copyright may be transferred without notice, after which this version may no longer be accessible.}
\doi{10.1109/ACCESS.2022.DOI}

\title{Systematic Literature Review on Cyber Situational Awareness Visualizations}
\author{\uppercase{Liuyue Jiang}\authorrefmark{\orcid{0000-0002-9419-1113}1,2},
\uppercase{Asangi Jayatilaka\authorrefmark{\orcid{0000-0003-2051-030X}1}, Mehwish Nasim\authorrefmark{\orcid{0000-0003-0683-9125}3,4}, Marthie Grobler\authorrefmark{\orcid{0000-0001-6933-0145}4}, Mansooreh Zahedi\authorrefmark{\orcid{0000-0001-6276-9956}5}, M. Ali Babar\authorrefmark{\orcid{0000-0001-9696-3626}1,2}}
}
\address[1]{CREST – the Centre for Research on Engineering Software Technologies, School of Computer Science, The University of Adelaide, Australia}
\address[2]{Cyber Security Cooperative Research Centre (CSCRC), Australia}
\address[3]{College of Science and Engineering, Flinders University, Adelaide, SA 5000 Australia}
\address[4]{CSIRO’s Data61, Melbourne, Australia}
\address [5] {School of Computing and Information Systems, The University of Melbourne, Australia}
\tfootnote{The work has been supported by the Cyber Security Research Centre Limited whose activities are partially funded by the Australian Government’s Cooperative Research Centres Programme.}

\markboth
{Jiang \headeretal: Systematic Literature Review on Cyber Situational Awareness Visualizations}
{Jiang \headeretal: Systematic Literature Review on Cyber Situational Awareness Visualizations}

\corresp{Corresponding author: Liuyue Jiang (e-mail: liuyue.jiang@adelaide.edu.au).}

\begin{abstract}
The dynamics of cyber threats are increasingly complex, making it more challenging than ever for organizations to obtain in-depth insights into their cyber security status. Therefore, organizations rely on Cyber Situational Awareness (CSA) to support them in better understanding the threats and associated impacts of cyber events. 
Due to the heterogeneity and complexity of cyber security data, often with multidimensional attributes, sophisticated visualization techniques are needed to achieve CSA. However, there have been no previous attempts to systematically review and analyze the scientific literature on CSA visualizations. In this paper, we systematically select and review 54 publications that discuss visualizations to support CSA. We extract data from these papers to identify key stakeholders, information types, data sources, and visualization techniques. Furthermore, we analyze the level of CSA supported by the visualizations, alongside examining the maturity of the visualizations, challenges, and practices related to CSA visualizations to prepare a full analysis of the current state of CSA in an organizational context. Our results reveal certain gaps in CSA visualizations. For instance, the largest focus is on operational-level staff, and there is a clear lack of visualizations targeting other types of stakeholders such as managers, higher-level decision makers, and non-expert users. Most papers focus on threat information visualization, and there is a dearth of papers that visualize impact information, response plans, and information shared within teams. Interestingly, we find that only a few studies proposed visualizations to facilitate up to the \textit{projection} level (i.e., the highest level of CSA), whereas most studies facilitated only the \textit{perception} level (i.e., the lowest level of CSA).  Most of the studies provide evidence of the proposed visualizations through toy examples and demonstrations, while only a few visualizations are employed in industrial practice.
Based on the results that highlight the important concerns in CSA visualizations, we recommend a list of future research directions. 

\end{abstract}


\titlepgskip=-15pt

\maketitle

\section{Introduction}

\emph{``The only truly secure system is one that is powered off, cast in a block of concrete and sealed in a lead-lined room with armed guards"}. These words by Gene Spafford illustrate the persistent vulnerability that networks and systems have in terms of cyber attacks, with cyber attacks increasing in sophistication and regularity. 
The outbreak of the COVID-19 pandemic has impacted every industry, particularly healthcare services, workers in remote areas, and the unemployed, who have all emerged to become new cyber attack targets \cite{cyberfact2021}.
A report published by IBM Security \cite{targetbreach2021} shows that the global average cost of a data breach in 2021 is estimated at US$\$$4.24 million, compared with US$\$$3.86 million in 2020 \cite{targetbreach}, with the latest statistics revealing that the average time for companies to identify a data breach in 2021 is $212$ days, up from $207$ days in 2020 \cite{targetbreach}. Particularly during the COVID-19 pandemic, many companies reported that they experienced the identification and containment of a data breach as taking longer. 
These statistics show that the number, depth, and breadth of incidents related to cyber attacks worldwide are increasing. 
Such incidents reinforce the need for better and faster mechanisms, tools, policies, risk management approaches, training, and technologies that can help safeguard the cyber environment of an organization. This all comes down to effective and efficient \emph{cyber security}. 

Cyber-related data is automatically generated at millisecond levels of resolution from diverse data sources and is often very voluminous. Furthermore, cyber attackers are increasingly applying sophisticated techniques in their attacks. As a result, implementing effective cyber security measures has become especially challenging. In this context, Situational Awareness (SA) has become paramount to facilitate correct and timely decision making to prevent or reduce the impact of cyber attacks. Situational (or situation) awareness is traditionally defined following the seminal work of Endsley \cite{endsley1995toward} as ``the perception of the element in the environment within a volume of time and space, the comprehension of their meaning, and the projection of their status in the near future''.  

Cyber data visualization can provide efficient and meaningful insights to overwhelming amounts of data, allowing decision makers to both explore and monitor the cyber status at various abstractions levels \cite{Franke2014}. Although various visualizations have been proposed to support CSA, there is no clear understanding of the different stakeholders for those visualizations, different types of information visualized, data sources employed, visualization techniques used, levels of CSA that can be achieved, and the maturity levels of the visualizations, challenges, and practices for CSA visualizations. 

Responding to this evident lack of investigation into an important topic, we systematically analyzed the literature on CSA visualizations. This Systematic Literature Review (SLR) enables both researchers and  CSA visualization designers to gain in-depth and holistic insights into the state-of-the-art CSA visualizations and offers support in transferring the research outcomes into industrial practice~\cite{kitchenham2004evidence}. Furthermore, the results can be used to identify limitations of the existing literature related to CSA visualizations and gaps that require further attention from the researchers.
The key contributions of this  systematic literature review  are listed below:

\begin{itemize}
  \item A synthesized body of research knowledge on CSA visualizations,  providing guidance for researchers and CSA visualization designers who want to better understand the topic.
  \item A comprehensive understanding of the different stakeholders of CSA  visualizations, different types of information visualized, data sources employed, and visualization techniques used.
    \item An analysis of the CSA levels that can be achieved through the proposed visualizations and the maturity of the proposed visualizations.
    \item An analysis of the challenges identified in designing and developing CSA visualizations and practices that have been reported to implement CSA visualizations successfully. 
    \item Identification of the potential gaps for future research highlights important and practical considerations for CSA visualizations that require further attention.
\end{itemize}

The rest of the paper is organized as follows: Section~\ref{sec:background} gives an overview of CSA and visualizations employed for achieving CSA. Section \ref{sec:methods} describes the methods that we used to conduct this SLR, including the review protocol. Section~\ref{sec:results} describes our results, including the demographic information and the quality assessment of the included studies, and addresses the research questions (RQs) through the analysis of selected studies. In Section~\ref{sec:discussion} we discuss  the key findings of this research and possible future directions. Section~\ref{threats} describes the threats to the research's validity. Finally Section~\ref{sec:conclusion} concludes the review.

\section{Background and related work} \label{sec:background}
This section discusses the background and related work with respect to  several important topics relevant to this SLR.  

\subsection{Situational Awareness}  \label{background:CSA} 
Situational  Awareness refers to the human cognitive capacity to analyze its environment and act accordingly. SA has been recognized as critical for successful decision making across a broad range of situations in various domains, including military command and control operations, health care, and air traffic control. 
SA is crucial for understanding and comprehending the implications of a situation, concluding, and making informed decisions about the future. 
It can be considered from two different aspects \cite{Franke2014}. 
The \textbf{technical aspect} of SA is concerned with collecting, compiling, processing, and fusing data. Here, information and data fusion is the most important concept considering aggregation and extraction of knowledge from various information sources to estimate current and predict future states.
The \textbf{cognitive aspect} of SA is concerned with a person's mental awareness in a given situation, specifically a person's capacity to comprehend the technical implications and draw conclusions to make informed decisions. 

Endley's model \cite{endsley1995toward} defines three SA levels  (see Figure~\ref{fig:sa}) that can be used to measure the extent to which a human decision maker is aware of a situation and whether they have reached a certain level of SA:

\Figure[][width=0.98\linewidth]{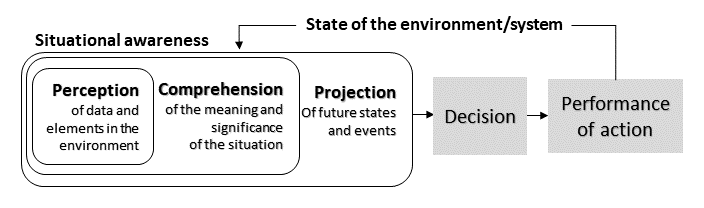}
{Three Levels of SA. Adapted from \cite{endsley1995toward}\label{fig:sa}.}

\begin{itemize}
\item{Perception (Level 1):} The lowest level of SA is associated with the user's perception of the status, attributes, and dynamics of the relevant elements of the environment.  

 \item{Comprehension (Level 2):} 
This involves comprehending or forming a synthesis of the situation based on the different elements in the perception level. This allows users to go beyond simply being aware of the elements in the environment to comprehending the situation and understanding the significance of those elements.

 \item{Projection (Level 3):} The highest level of SA is associated with the ability to predict future states or events of the elements of the environment. The accuracy of the prediction highly depends on the accuracy of SA Level 1 and Level 2. 
\end{itemize}
It is important to note that the proposed levels of SA represent ascending levels of awareness and not linear stages~\cite{endsley2015situation}. 
By following this process, the user can rationalize the situation at hand, enabling decision making and action.  The person who comprehends and understands the meaning of the current situation will possess greater situational awareness than a person who reads the data without understanding its meaning. Similarly, someone who can predict probable future events and states will better understand the situation than someone who is unable to do so.

\subsection{Cyber Situational Awareness}
Given the progressiveness and usefulness of SA research, it is increasingly applied to cyberspace~\cite{Franke2014}.   Hence, CSA can be considered an extension or a subset of traditional SA to cyberspace.

A systematic literature review on CSA conducted by Franke et al. \cite{Franke2014} describes and discusses peer-reviewed literature on this topic from the perspective of both national cyber strategies and science.  SA requires adequate  knowledge about the organization's current and past cyber activities to effectively detect, identify, and respond to various threats and attacks within the cyber security domain. CSA provides holistic and specific information related to cyber threats and vulnerabilities, allowing organizations to swiftly identify, process, and comprehend information.  
Such suspicious and interesting activities can be diverse and might range from low-level network sniffing to activities obtained by external data sources such as social media. In turn, CSA helps organizations understand their current and future risk situation and position in terms of their protection mechanisms.

In line with the three levels of SA, CSA is concerned with developing the ability to recognize the current state of assets and the cyber threat situations (\textit{perception}), the ability to comprehend the meaning of the cyber threat situation and assess the impact of the threats  (\textit{comprehension}), and the ability to project the future state of threats or actions (\textit{projection}).

Current CSA research mainly focuses on three aspects: data collection \cite{8915712, 7317717, 10.1145/3230833.3232798}, data processing and analysis \cite{vinayakumar2019deep, hariharan_camlpad_2020,alrashdi2019ad}, and  data visualization \cite{carroll2019makes, pid154}. Newer models and frameworks have been proposed to achieve CSA, such as cyber-specific Common Operating Pictures (COPs)~\cite{conti2013towards}. COP has historically been a military term used to describe a command and control solution that aggregates important operational information into a single picture, \textit{\say{a single identical display of relevant information shared by more than one command that facilitates collaborative planning and assists all levels of decision makers to achieve situational awareness}}. Conti et al. \cite{conti2013towards} clearly articulated the roles of humans and machines in a Cyber Common Operating Picture (CCOP) for achieving CSA. They argued that CCOPs should be designed to consider the tasks better suited to human cognitive capabilities, and those can be automated and processed at high speed by machines.

Advanced sophisticated data analytic techniques are often used to process and analyze complex cyber information in real-time and offline to provide CSA.   However, due to the volume and complexity of cyber data and attacks,  powerful machine learning techniques alone are insufficient to achieve CSA \cite{fink2009visualizing, conti2013towards}. It is important to effectively link technical aspects with cognitive aspects in cyber security to achieve complete CSA. To this end, effective data visualization is imperative; visualizations allow users to explore and analyze large amounts of data and quickly identify trends and unexpected events, enabling swift decision making and action \cite{Franke2014, fink2009visualizing, pid107}.

\subsection{Cyber Situational Awareness Visualizations}
Although we observed an increasing numbers of papers in the literature around the   topic of CSA visualizations, we did not find any existing SLR or systematic mapping study focused on the visualizations aimed at CSA. However, there have been several existing reviews on visualizations for specific security areas. This section compares these existing reviews and discusses the gaps and novelty of this SLR.

A number of studies look at visualizations related to network \cite{shiravi2011survey, zhang2017survey, guimaraes2015survey} and malware analysis \cite{wagner2015survey}.  For instance, Shirave et al. \cite{shiravi2011survey} present an SLR on network security visualizations. 
The authors identified five network security visualization classes, including host server monitoring, internal/external monitoring, port activity, attack patterns, and routing behavior. In another study, Guimaraes et al. \cite{guimaraes2015survey} present an SLR  of information visualization for network and service management. They identified several well-explored topics on network and service management regarding the use of information visualization, including IP networks, monitoring, and measurement. They also analyzed the visualization techniques and tasks/interactions in  network and service management information visualizations. Their results revealed that standard 2D/3D displays are the most commonly used visualization technique in network and service management visualizations. They also point out a number of future research directions for information visualizations for network and service management, specifically \textit{IoT}, \textit{Big data}, \textit{Cloud computing}, \textit{SDN}, and \textit{Human-centered evaluation}.  

Wagner et al. \cite{wagner2015survey} provide a survey of visualization systems for malware analysis. They categorized the literature based on a general approach to data processing and visualization using a malware visualization taxonomy. They also categorized the literature by their input files and formats, the visualization techniques utilized, the representation space and the mapping to time, certain temporal aspects, their interactive capabilities, and the different types of available user actions. 

Staheli et al.~\cite{staheli2014visualization} provide a survey of visualization evaluations for cyber security. The authors identify the most common evaluation types for  security applications and discuss future directions. Franke et al. \cite{Franke2014} conducted an SLR that specifically focused on CSA. Their survey is broad and includes publications that are not related to visualization. They focus on various topics, including introductory literature on CSA and SA in industrial control systems, emergency management, SA architectures, algorithms, and visualizations.  In terms of visualizations, Franke et al. \cite{Franke2014} specifically highlight the need for going beyond technical aspects of the visualizations to obtain a more comprehensive understanding of the relationship between CSA levels (i.e., mental states) and the CSA visualizations. 

In summary, there are several shortcomings in the existing literature reviews. 
Most of the reviews mentioned above have not been carried out considering  CSA or only consider specific areas related to cyber security visualizations (e.g., network analysis or malware analysis).
Therefore, the existing literature reviews do not provide an overall view of CSA visualizations.
Furthermore, existing literature does not consider or describe important aspects such as the level of SA that could be reached (i.e., mental states) using the visualization, diversity of stakeholders,  types of information visualized, and challenges and practices for CSA visualizations. 
Therefore, in this research, we conduct an SLR to obtain a complete view of the literature on visualizations targeting CSA while considering the aspects mentioned above that are important to the CSA domain, narrowing the existing knowledge gap in this field. 

\section{Methodology} \label{sec:methods}

\Figure[t!](topskip=0pt, botskip=0pt, midskip=0pt) [width=0.9\textwidth] {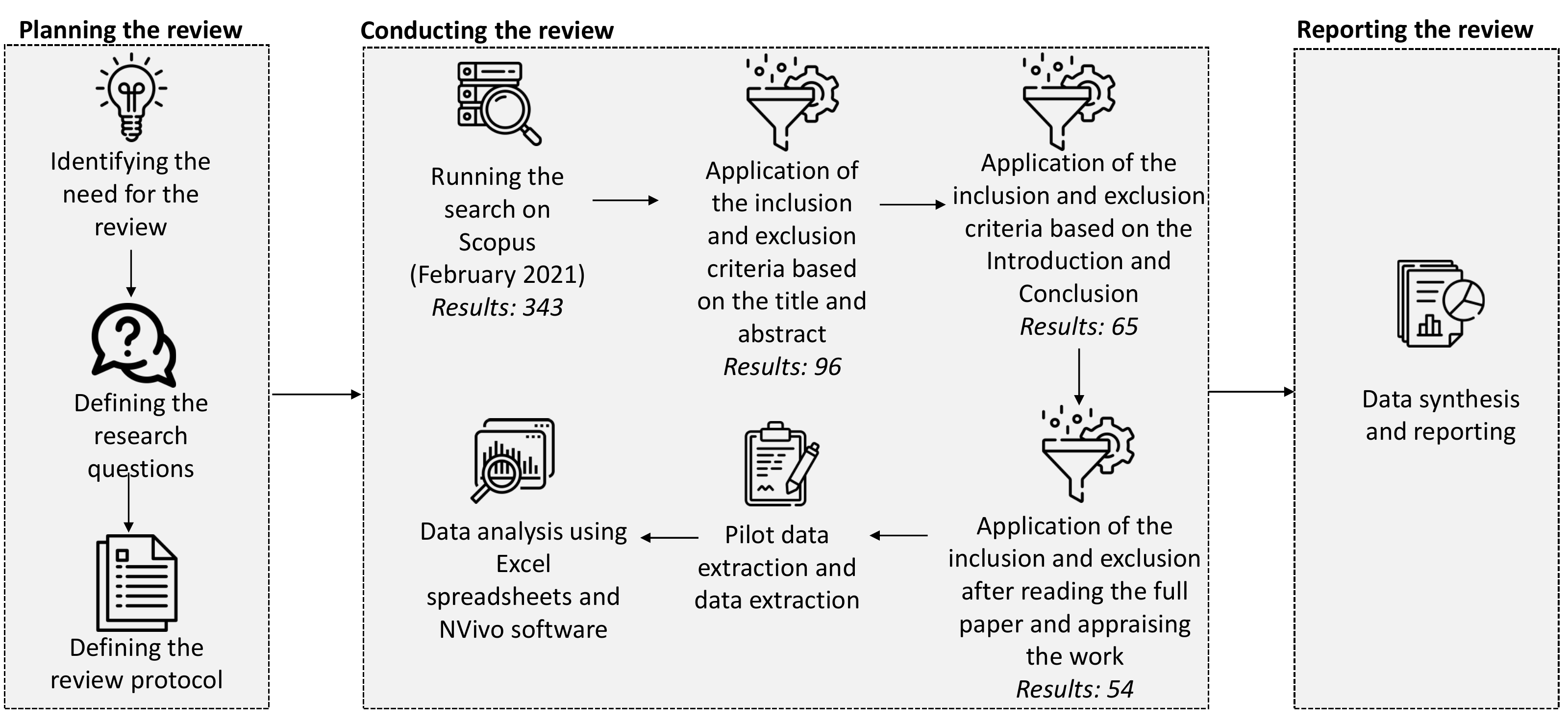}
{Methodology\label{fig:studyselection}}

The research methods in an SLR  provide a well-defined process for identifying, analyzing, and interpreting  literature relevant to a particular set of research questions (RQs). We followed the three-phased guidelines published by Kitchenham and Charters \cite{kitchenham2007guidelines}: defining a review protocol, conducting the review, and reporting the review. We describe the main steps of this SLR in the following subsections, detailing the process illustrated in Figure~\ref{fig:studyselection}.

\subsection{Research questions}
This SLR focuses on providing an extensive overview and analysis of the existing literature on CSA. Therefore, we formulated five RQs to guide this SLR.  Table \ref{tab:RQs} presents the RQs, along with their motivation. 

\begin{table}[]\caption{\label{tab:RQs} Research Questions}
\begin{tabularx}{0.48\textwidth}{lp{0.15\textwidth}p{0.21\textwidth}}
\toprule
Number & Question & Motivation \\
\midrule
RQ1  &   Who are the stakeholders that use and benefit from CSA visualizations?   &  To understand the types of people who are intended to benefit from the proposed CSA visualizations.  \\
RQ2 & 
What types of information are visualized, which data sources are used, and how is the cyber information visualized?
& To understand the different types of information presented in CSA visualizations, data sources used, and visualization techniques and task interactions employed to visualize CSA data.  \\
RQ3 & What level of CSA is facilitated by the visualizations?  & 	
   To understand the level (i.e., \textit{perception}, \textit{comprehension} and \textit{projection})  supported by visualizations.\\
RQ4&  What is the maturity of the proposed visualizations that facilitate CSA?   & To help researchers assess the maturity of the CSA visualizations. \\
RQ5   &  What are the reported challenges in employing visualizations to facilitate CSA? &  To identify the challenges for designing and developing CSA visualizations reported in the literature.\\
RQ6 &   What practices have been reported to implement CSA visualizations successfully? & To understand the good practices, guidelines, lessons learned, and shared experiences needed to implement CSA visualizations.\\

\bottomrule
\end{tabularx}
\end{table}

Answering these RQs will provide an in-depth understanding of the stakeholder of the CSA visualization (RQ1), the types of information visualized, the data sources used, and how the cyber information is visualized (RQ2), the level of CSA is facilitated by the visualization (RQ3), CSA visualization maturity (RQ4), challenges for CSA visualizations (RQ5), and practices for supporting effective CSA visualization (RQ6). In addition, the findings will enable researchers to obtain an in-depth overview of this topic, identify limitations and gaps,  and potential future directions. 

\subsection{Search strategy} \label{serachstratergy}
This subsection discusses the search terms and data sources used in this SLR.
We used the guide presented by \cite{kitchenham2010systematic} to develop the search string of this study iteratively. First, the base keywords used as search terms were constructed by considering the three aspects related to the SLR topic: cyber, situational awareness, and visualizations.  Then, we systematically modified the search string by adding a set of alternative search terms. These alternative search terms were obtained by considering researchers' knowledge and experience, synonyms, and key terms used in the existing related research papers.

\begin{itemize}
    \item \textbf{Cyber} --	cyber* 
    \item \textbf{Visualizations} -- visual* OR dashboard OR dash board OR dash-board OR picture OR diagram OR graphic OR video OR  image OR audio OR multimedia OR multi media OR  multi-media
    \item \textbf{Situational awareness} -- situational aware* OR situational-aware* OR situation aware* OR common operating picture OR common operational picture OR CCOP
\end{itemize}

The identified search terms were combined into the final search string using Boolean AND and OR operators.
 We conducted several pilot searches to identify the  best search string. We also  verified the inclusion of well-known primary studies when finalizing the search terms.  The final search string combined each base keyword category with an AND operation:
 
\textit{cyber* AND (visual* OR dashboard OR  dash board OR dash-board OR picture OR diagram OR graphic OR video OR image OR audio OR multimedia OR multi media OR multi-media) AND (situational aware* OR situational-aware* OR situation aware* OR common operating picture OR common operational picture OR CCOP)}. 

Previous researchers \cite{kitchenham2010systematic, shahin2020architectural} have shown that Scopus indexes a large number of journals and conference papers indexed by other search engines, including ACM Digital Library, IEEE Xplore, Science Direct, Wiley Online Library, and SpringerLink.  Furthermore, digital libraries such as  SpringerLink and  Wiley Online Library place several restrictions on the meta-data of the published studies in large-scale searches. 
The search string also needs to be modified for each digital library, otherwise it would result in errors being introduced.
As such, we used the Scopus search engine to find potentially relevant papers. Scopus enabled us to use one search string while retrieving the most relevant studies. The search terms were matched with the title, abstract, and keywords of papers in Scopus. The search conducted in February 2021 resulted in 343 papers that matched the search string.

\begin{table}[]\caption{\label{tab:inclusionExclusion} Inclusion (I) and exclusion (E) criteria}
\begin{tabularx}{0.48\textwidth}{p{0.07\textwidth}p{0.36\textwidth}}
\toprule
Inclusion or Exclusion & Criteria \\
\midrule
I1 & The paper should introduce a CSA visualization with design or implementation.\\
I1  &  The paper should be peer-reviewed.  \\
I2 & The paper should be published in the English language.  \\
E1 & Any editorials, position papers, keynotes, reviews, tutorial summaries, and panel discussions are excluded.\\
E2 & 	If a conference paper and a journal paper duplicate the same work, the conference paper will be excluded, and the journal paper will be retained.\\
E3 &	Short papers of fewer than five pages are excluded. \\
E4 &	Papers for which the full text is not available at the time of the study are excluded. \\

\bottomrule

\end{tabularx}
\end{table}

\subsection{Study selection} \label{studyselection}
Three authors applied the inclusion and exclusion criteria detailed in Table~\ref{tab:inclusionExclusion} to systematically select the final set of papers included in this SLR.  All the authors discussed the criteria and agreed upon them before the study selection phase. We refined the inclusion and exclusion criteria in several iterations  to accurately  classify the papers.  One of the critical selection criteria is  that the paper should introduce a visualization for CSA with a design or implementation (I1). In the meantime, we decided  not  to include any short papers (E3) because they only presented concepts or ideas. They lack  well-defined  visualizations, and most importantly, they did not provide sufficient and relevant evidence to answer the defined RQs. 

By applying the inclusion and exclusion criteria to the papers' titles and abstracts, the number of papers was reduced to $96$. The inclusion and exclusion criteria were then applied to the introduction and conclusion of the remaining papers, resulting in the further exclusion of $31$ papers. The majority of the papers excluded at this point resulted from the papers not specifically addressing visualizations for CSA. For example, we excluded papers that mainly address physical infrastructure in the smart-grid industry. We read the full text of the remaining $65$ papers in the last stage and included only $54$ in the final set. For example, we excluded the papers claiming CSA visualization in the abstract and introduction but which do not have proper visualization design or implementation. The three authors' disagreements during the study selection were discussed with the other authors in detail and resolved before moving on to the data extraction.

\subsection{Data extraction} \label{dataextraction}

Data extraction was performed by three authors, following the guidelines set out by Kitchenham et al. \cite{kitchenham2007guidelines}, where multiple researchers review different primary studies due to time or resource constraints. This process recommends a method of checking to ensure that researchers extract data consistently. A pre-defined data extraction form (see Table~\ref{tab:dataextraction}) was used to extract data from the selected primary studies. 
When extracting data, we considered a single visualization as a region in a user interface with a clear visual boundary where information is displayed as a group. Before the data extraction, we conducted a pilot data extraction and compared the results of a selected random sample of primary studies to ensure the data extraction form could capture all the required information in the best possible summarised version.  Any disagreements were discussed in detail and resolved before moving on to the data extraction from all the papers.

\begin{table}[]\caption{\label{tab:dataextraction} Data extraction form}
\begin{tabularx}{0.48\textwidth}{lp{0.08\textwidth}lp{0.19\textwidth}}
\toprule
Item  & Question & Related RQ & Description \\
\midrule
D1&	Authors&	Demographics& --	\\ 
D2&	Year&	Demographics& --	\\ 
D3&	Publication type&	Demographics& --	\\ 
D4&	Publication venue&	Demographics& --	\\ 
D5&	Stakeholders&	RQ1 &	Types of people intended to benefit from the proposed CSA visualization \\ 
D6&	Information visualized &	RQ2 & Types of information that are visualized \\
D7&	Data sources & RQ2 &  Data sources used for visualization	\\ 
D8& Visualization techniques & 	RQ2  & Visualization techniques employed \\ 
D9&	Tasks or interactions &	RQ2 &  The ways in which a user can interact with the visualizations \\ 
D10&	Level of CSA  &	RQ3 & Level of CSA facilitated through the visualizations\\ 
D11&	Maturity of visualizations &	RQ4 &  Assessment of  the reported evidence \\ D12&	Challenges reported &	RQ5 &	Challenges and barriers that have been reported to design, implement and adopt CSA visualizations\\ 
D13&	Practices&	RQ6 &	Lessons learned, good practices, and authors' experiences in successfully implementing  CSA visualizations\\ 
\bottomrule
\end{tabularx}
\end{table}

\subsection{Data analysis and synthesis} \label{sec:dataanalysis}

The demographic and contextual set of data items (D1 to D4 in Table~\ref{tab:dataextraction}) were analyzed by employing descriptive statistics. Other extracted data (D5 - D13) to answer the RQs were analyzed using thematic analysis or existing taxonomies. Thematic analysis was used where taxonomies were not available to analyze the collected data.  We describe in detail how the data was analyzed below. 

\subsubsection{Thematic analysis for qualitative data analysis}
The data extracted for D5, D6, D7, D12, and D13 were analyzed using the thematic analysis technique. Thematic data analysis is a widely used qualitative data analysis method. We used the steps proposed by Braun and Clarke to thematically analyze the qualitative data collected~\cite{braun2012thematic}.  First, we familiarized ourselves with the extracted data by carefully reading each element. After familiarizing ourselves with the data, the data were saved to the NVivo data analysis tool for further analysis.  Based on the principles of thematic analysis, we then performed open coding. This involved  breaking the data into smaller components to generate the initial codes.  The key points of the data were summarized using codes (i.e., a phrase) of three-five words. Next,  codes were  grouped and  assigned to potential themes. This was an iterative process as it was important to revise and merge codes based on their similarities.

\subsubsection{Use of existing taxonomies for analyzing the extracted data} \label{dataanalysis_taxanomies}
To analyze the data extracted for D8, D9, D10, and D11, we utilized existing taxonomies.  
We observed a range of taxonomies proposed in the information visualization field to analyze data collected for visualization techniques (D8). However, some of these were too specific or unrelated to our purpose. For example, the researchers in \cite{borkin2013makes} propose a taxonomy specifically for static (i.e., non-interactive) visualizations, whilst other specific taxonomies have been proposed for dynamic graph visualizations~\cite{beck2017taxonomy, lee2006task} and treemap visualizations~\cite{scheibel2020taxonomy}. 

The taxonomy proposed by Guimarães et al. \cite{guimaraes2015survey} is closely related to our work.  They merged two taxonomies  %\cite{keim2002information, shneiderman1996eyes} 
to achieve the framework needed for an adequate general classification of \textit{visualization techniques} and  \textit{tasks or interactions} for end-users. For the first criterion (i.e., visualization techniques), the researchers in \cite{guimaraes2015survey} used the  “Information Visualization and Data Mining” taxonomy proposed in \cite{keim2002information}.  These taxonomies are widely accepted and referenced by the visualization community. 
Based on the visualization technique taxonomy proposed by Guimarães  et al. \cite{guimaraes2015survey}, it is possible to divide the  techniques used in the visualizations into five generalized categories: i)~~\textit{standard 2D/3D displays}, ii)~\textit{geometrically transformed displays}, iii)~~\textit{iconic displays}, iv)~~\textit{dense pixel displays}, v)~~\textit{stacked pixel displays}. In addition to these categories, we added four more categories: \textit{geographical displays}, ~\textit{immersive environments}, \textit{single value displays}, and \textit{tables/text summaries}, to capture the visualization techniques we observed in our papers. Detailed descriptions of these categories are given in Section~\ref{ss:visualization techniques}.

To analyze the \textit{tasks/interactions} (D9), Guimarães et al. \cite{guimaraes2015survey} merged the taxonomies proposed by Keim \cite{keim2002information} and Shneiderman \cite{shneiderman1996eyes}, and they added a new task and interaction technique called \textit{move/rotate}. The resulting taxonomy had nine categories: i)~\textit{overview}; ii)~\textit{zooming}; iii)~\textit{filtering}; iv)~\textit{details on demand}; v)~\textit{history}; vi)~\textit{relate}; vii)~\textit{extract/share}; viii)~\textit{move/rotate}; and ix)~\textit{linking and brushing}.  We added a new  task/interaction  called \textit{customization}  to capture information on user interactions related to customization of visualizations. It is important to note that the  \textit{overview} category is concerned with the ability to gain an overview of the entire data collection %. We  observed that an \textit{overview} of the entire data collection can be gained 
using other \textit{tasks/interactions}, such as  \textit{zooming} and \textit{filtering}. Hence to remove the duplication of information, we  did not use the \textit{tasks/interactions} called \textit{overview} in this study. Detailed descriptions of the \textit{tasks/interactions}  used in this study are given in Section~\ref{ss:interaction techniques}.

Using data collected in D10, we explain how the visualizations support SA. Here we mapped each visualization to the three levels of SA defined by Endsley~\cite{endsley1995toward}. Data collected for D11 are used to explain the maturity of the proposed visualizations that facilitate CSA. We have used a six-level hierarchy proposed in \cite{alves2010requirements} to describe the visualization maturity. The details of this hierarchy proposed in \cite{alves2010requirements} are given in Section~\ref{maturity}.

\subsection{Quality assessment} \label{qualityassestment}
The 54 primary studies were evaluated by the same three authors who performed the data extraction. The quality assessment was performed against  the set of quality assessment questions listed in Table~\ref{tab:qualityresults} (adopted from \cite{shahin2014systematic, dybaa2008empirical}). Each question was answered during the data extraction process according to a ratio scale `Yes', `No', or `Partially'. The answers for each study show the quality of a selected study and the credibility of the study’s results. Previous studies highlight that the quality assessment result of the included studies can reveal the potential limitations of the current research and guide future research in the field \cite{dybaa2008empirical,kitchenham2007guidelines}.
Similar to \cite{shahin2014systematic}, the quality assessment was not used for study selection but was employed for validating the results of the selected studies. 

\begin{table}[]\caption{\label{tab:qualityresults}Study quality assessment results}
\resizebox{0.48\textwidth}{!}{%
\begin{tabular}{p{0.18\textwidth} l l l}
\toprule
\textbf{Study quality assessment question} & \textbf{Yes} & \textbf{Partially} & \textbf{No} \\ \midrule
Is there a rationale for why the study was undertaken? & 54 (100.0\%) & 0 (0.0\%) & 0 (0.0\%) \\ 
Is there an adequate description of the context? & 48 (88.9\%) & 6 (11.1\%) & 0 (0.0\%) \\
Is there a justification and description of the research design? & 37 (68.5\%) & 15 (27.8\%)  & 2 (3.7\%) \\ 
Has the study an adequate description of the technique for visualization? &  38 (70.4\%)& 15 (27.8\%)&  1 (1.9\%)   \\ 
Is there a clear statement of the findings?  &  33 (61.1\%) & 16 (29.6\%) & 5 (9.3\%) \\ 
Do the researchers critically examine their potential bias and influence on the study?   & 5 (9.3\%)  & 17 (31.5\%) &   32 (59.3\%)\\
%7 & \begin{tabular}[c]{@{}l@{}} Do the authors discuss the credibility of their findings? \end{tabular}  & 2 (6.1\%) & 5 (15.2\%) & 26 (78.8\%)  \\ \hline
Are the limitations of the study discussed explicitly?  & 6 (11.1\%) & 9 (16.7\%) & 39 (72.2\%)  \\ \hline
\end{tabular}}
\end{table}

\section{Results} \label{sec:results}
This section reports the synthesis and analysis results of the data extracted from the 54 primary studies to answer the research questions. 

\subsection{Demographics} \label{sec:demographics}

\begin{figure} 
\centering
\includegraphics[width=0.48\textwidth]{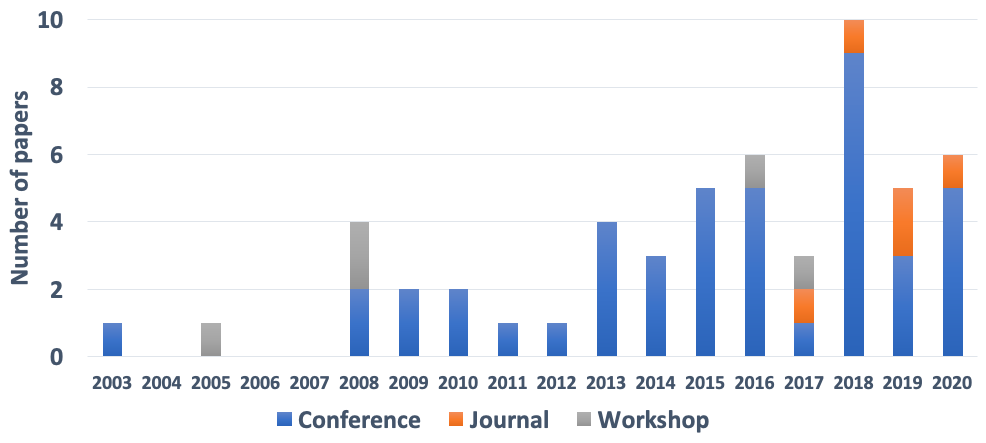}
\caption{Distribution of papers in years and venue types}
\label{fig:years}
\end{figure}

Our dataset comprises papers published between 2003 and 2020. Only eight papers in our dataset were published before 2010, with the remaining 46 papers ($85.2\%$) published in or after 2010. Of those papers, about 44.4\% (24 papers) were published in or after 2017. It shows that CSA has only started to gain popularity in the last decade. The distribution is shown in Figure~\ref{fig:years}.
Most selected studies were published in conferences (44 studies, 81.5\%). Only five studies (9.3\%) were published in workshops. The remaining five studies (9.3\%) were published in journals. We found that the \textit{International Symposium on Visualization for Cyber Security  (VizSec)} is a  popular venue for publishing work on CSA visualizations as they have published 13.0\% (7 studies) of the selected studies. The \textit{International Conference on Cyber Situational Awareness, Data Analytics and Assessment (CyberSA)} has three publications (5.6\%), and the \textit{International Conference on Big Data} has two publications (3.7\%). 
Most other venues only show one paper. The selected studies were generally published in venues targeted at security, visualization, big data, and general software engineering. This finding demonstrates that this research topic has been broadly considered by different research interests.

\subsection{Quality assessment results}
\label{qualityresults} 

Table~\ref{tab:qualityresults} illustrates the quality assessment results of the 54 publications selected. As shown in Table~\ref{tab:qualityresults}, all studies state the rationale for the conducted study (Q1). Q2 was answered positively by most studies (88.9\%), which means the reviewed studies have an adequate description of the context in which the research was carried out. Concerning Q3, 37 out of 54 studies (68.5\%) provided adequate descriptions of the research design (Q3). The answers to Q4 and Q5 reflect the accuracy of the data extraction results. 38 out of 54 studies (70.4\%) described their proposed visualization techniques adequately, and 15 studies (27.8\%) addressed these techniques to some extent.  61.1\% of studies have a clear statement of their findings. Q6's majority (59.3\%) “No” responses show that the researchers did not critically examine their bias and influence on the study's outcomes. The majority of the studies (72.2\%) did not discuss any limitations or drawbacks. 

\subsection{RQ1: Who are the stakeholders that use and benefit from CSA visualizations? } \label{sec:stakeholders}

\vspace{1em}
\begin{mdframed}
     \begin{itemize}
  \item CSA visualizations found in the primary studies mainly target three types of stakeholders: i) operational-level staff; ii) managers and senior-level decision makers; and iii) non-expert users.
  \item Most visualizations focus on operation-level staff. However, only a few studies focus on managers, senior-level decision-makers, and non-expert users.
\end{itemize}
\end{mdframed}

This section presents the findings for RQ1 and describes the various stakeholders of CSA visualizations. 
The data extracted for this section correspond to item D5 in Table~\ref{tab:dataextraction}.
In our selected primary studies, we found three main categories of stakeholders, noting that several papers targeted multiple stakeholders. These three categories of stakeholders are described below.

    \textbf{Operational-level staff:} We found that the majority of selected primary studies targeted the  \textit{operational-level staff} and focused on facilitating their day-to-day business ($64.8\%$). Among these papers, some visualizations targeted network analysts.  For example, researchers in [P38] propose a scalable platform for large-scale networks to process and visualize data in real time. Furthermore, researchers in  [P47] propose an ensemble visualization approach to improve network security analysis. Another set of papers focuses on CSA visualizations targeting risk analysts and security analysts. For example, researchers in [P54] propose multiple views that allow security analysts to analyze the event history, asset relationships, and plausible future events to identify the best course of action.
     
     \textbf{Managers and higher-level decision makers:} With cyber attacks becoming more frequent, sophisticated, targeted, and widespread, cyber security decision makers need to make quick critical decisions to contain and mitigate cyber attacks. Several studies have focused on CSA visualizations to assist \textit{managers and higher-level decision makers} ($35.2\%$) in assessing risks, allocating resources, and altering the state of operations of the organization in response to the real and potential security risks. For example,  researchers in [P5] propose a Cyber COP that facilitates commanders' decision making process by recognizing the current state of assets and the cyber threat situation, the impact of cyber attacks on the mission related to assets, and future threat scenarios. In [P49], the researchers demonstrate how composite visual data structures and their synthesis can reduce or illuminate the direction of cyber security policies. 
    
     \textbf{Non-expert users:} Two primary studies (3.7\%) focus on CSA visualizations tailored to \textit{non-expert users}. In particular, these two papers [P1, P44] propose CSA visualizations to enable \textit{non-expert users} to actively monitor and observe their activity for greater online awareness. While [P44] focuses on 2D visual analytics interfaces,  [P1] engages in 3-dimensional visualizations for home networking monitoring.

\begin{table}[]
\caption{\label{tab:stakeholders}Stakeholders}
\begin{tabular}{p{0.09\textwidth} p{0.28\textwidth} p{0.05\textwidth}}
\hline
\toprule
 Stakeholder &  Papers  & Count \\
\midrule

 Operational level staff &  
 [P3, P4, P6, P7, P9, P11, P12, P13, P14, P16, P19, P20, P22, P24, P26, P27, P28, P30, P32, P33, P34, P35, P37, P38, P39, P40, P41, P42, P45, P47, P48, P50, P51, P52, P54]
 
 &  35 \\ 
 Managers and higher level decision makers & [P2, P5, P10, P12, P13, P16, P21, P22, P23, P24, P29, P31, P32, P36, P37, P43, P46, P49, P53]  &  19 \\ 
 
 Non-expert users  & [P1, P44]  &  2 \\ \bottomrule
 
\end{tabular}

\end{table}

\subsection{RQ2: What are the types of information visualized, data sources used, and how is the cyber information visualized? } \label{ss:usecases}

\vspace{1em}
\begin{mdframed}
     \begin{itemize}
  \item Various types of information are represented  through  CSA visualizations. \textit{Threat information} is the most common type of such information. However, only a few studies consider \textit{impact information}, \textit{response plans}, and \textit{shared information}. 
  
 \item Often multiple data sources are utilized together in CSA visualizations.  The most frequent data sources are \textit{asset identification systems} and  \textit{logs}. The \textit{external data sources} and \textit{human input and organizational information} are the comparatively less common  data sources for CSA visualizations.
 
 \item   \textit{Iconic displays} and \textit{geometrically transformed displays} are the prominent types of visualization techniques employed in CSA visualizations. On the other hand, \textit{immersive environments} are very rarely used in CSA visualizations.
 
 \item Only a few interaction techniques are used in CSA visualizations frequently. These are \textit{zooming}, \textit{filtering}, and \textit{details on demand}. Other interaction techniques such as \textit{relate}, \textit{extract/share}, \textit{move/rotate}, \textit{linking and brushing}, and \textit{customization} are very rarely employed in CSA visualizations.
  
\end{itemize}
\end{mdframed}

This section presents the findings for RQ2. In particular, we discuss different types of information visualized in the CSA visualizations (see Section~\ref{sec:whatInfo}), data sources used (see Section~\ref{sec:datasources}), and how the cyber information is visualized (see Section~\ref{visualizations}).

\begin{table}[]
\caption{\label{tab:whatinfo}Information types}
\begin{tabular}{p{0.1\textwidth} p{0.26\textwidth} p{0.04\textwidth}}
\hline
\toprule
 Information types &  Papers  & Count \\
\midrule

 Assets &  [P1, P5, P8, P13, P20, P22, P30, P31, P33, P34, P37, P39, P40, P41, P43, P45, P46, P51, P53, P54]  &  20 \\ 
 History and trends &  [P2, P4, P7, P16, P21, P24, P25, P27, P30, P34, P37, P38, P44, P47, P48, P49, P50, P53, P54] &  19 \\ 
 Impact information  &  [P2, P8, P9, P13, P21, P24, P33, P43, P54] &  9 \\
  Response plans   &  [P2, P8, P16, P21, P28, P31, P33, P36, P39, P51] &  10\\
   Shared information   &  [P3, P10, P12, P31, P37, P51] & 6 \\
      Network information   &  [P1, P5, P11, P12, P13, P14, P15, P24, P29, P34, P35, P36, P38, P40, P42, P44, P47, P49, P50, P51, P52]    & 21 \\
      Risk information   &  [P2, P4, P6, P9, P10, P12, P13, P14, P16, P17, P19, P21, P23, P25, P31, P32, P33, P36, P37]    &  19 \\
          Threat information   &  [P2, P3, P5, P7, P8, P10, P13, P16, P18, P19, P20, P21, P22, P23, P25, P26, P28, P30, P31, P32, P33, P34, P35, P37, P39, P41, P45, P46, P47, P48]   & 30 \\
 \bottomrule
 
\end{tabular}
\end{table}

\subsubsection{What information is visualized} \label{sec:whatInfo}
This section presents the types of information visualized in our primary studies.  The data extracted for this section corresponds to item D6 in Table~\ref{tab:dataextraction}. Our thematic analysis resulted in the identification of eight types of information, as shown in  Table~\ref{tab:whatinfo}.   A single paper may visualize multiple types of information hence may have repeated entries in the table.

    \textbf{Assets:} An asset in the context of cyber security could be any data, device, or other components of an organization’s systems that are valuable, mainly because it contains sensitive data or can be used to access such information. Therefore, a clear understanding of the assets-related information is vital to CSA. We found 20 papers (37.0\%) in our SLR that visualized asset information. Among our primary papers, it was common to employ map views to visualize organizational cyber assets, geographic locations to which the target assets belong, and the relationship between those assets [P5, P8, P31]. Apart from this, cyber capabilities critical to the mission, network state in terms of assets, assets, and their relationship with cyber incidents were visualized in our primary studies. 

    \textbf{History and trends:} Analyzing the history and trend information allows users to easily contextualize the current cyber security status. Furthermore, understanding the trends and patterns allows users to make predictions with some certainty. In our selected papers, we found 19 papers (35.2\%) that visualized history and trend information. It involves historical data related to attacking behavior, temporal information related to cyber security incidents, and trends in overall organizational performance. For example, we observed several papers provide the temporal context of cyber events to the users by displaying relevant data that happened before an event occurred [P4, P25]. Researchers in [P21] propose novel circle-based cyber security metric display visualizations capable of displaying history information along with the current metric values. However, only a few studies visualized history or trends for overall organizational performance. For example, researchers in [P13] provide views for high-level management to analyze the history and trends related to the impact of compromised network nodes and the cost of corrective actions.

    \textbf{Impact information:}  Understanding the impact or consequences of successful or potential cyber security events is crucial in identifying how to respond to those incidents or possible attacks. A limited number of papers provide various visualizations to support this (16.7\%). For example, in [P13], the visualization uses the concept of area corruption to convey visually the impact of a compromised device on its supported process. Each compromised device will produce a hole in the area proportional to its operational impact score value. In [P2], researchers propose a proactive environment that shows the maximum impact or risk for the business devices. 
   
     \textbf{Response plans:} Some papers (18.5\%) provide visualizations to assist users in determining the response plans for cyber incidents are grouped under this category. For a given situation, there can be multiple response methods. The visualizations in the selected set of papers assist users in either identifying these response methods or selecting the most suitable response plan by analyzing their costs and benefits. For example, researchers in [P5] propose visualizations that allow doing ``what if" projections to explain to commanders the cyber side of the different “courses of action” (CoAs) that are proposed to the commanders by their staff. In another example, response plans are presented to users in various dimensions such as risk mitigation, return on responsible investment, and impact [P2].

     \textbf{Shared information:}
     Achieving complete situation awareness requires members of different teams and different organizational positions, working across different work shifts to collaborate and share information.  
     In our primary papers, we observed a limited number of studies (11.1\%) that include visualizations to support communication and collaboration among different team members. These visualizations consist of information related to observations and hypotheses performed or insights gained by the analysts. They also include analyst movements for the coordinators, email communication with the team, and communication workflows. For example,  researchers in [P3] focus on a mind mapping system for supporting collaborative cyber security analysis, and researchers in [P37] propose visualizations to show shared incident reports and as well as to facilitate the coordination of incident responses and defenses among the multiple stakeholders. 
    
     \textbf{Network information:} Several papers in our data set visualize various network-related information~(38.9\%). The visualized information in this category includes network data, network topology, network reports, and network communication. For example, in [P38], both streaming and archived network flow data is visualized in real-time to support network activity monitoring, identify network attacks and compromised hosts, and anomaly detection. In another example, visualizations were proposed to analyze firewall log events [P50].
   
     \textbf{Risk evaluation:} We observed several primary papers (35.2\%) in this SLR that visualize information that allows the user to assess the risks related to possible attacks and threats. Risk evaluation information can take on various forms. For example, known vulnerabilities on critical assets can be related to security alerts for risk evaluation [P33], changes in risk levels [P6, P13], possible attack paths [P28], suspicious or known malicious IP addresses [P4], classification and distribution of cyber security events [P19, P21, P23, P25], and attacker capacities [P16].  

     \textbf{Threat information:} Threat information is the most highly sought piece of information in our primary studies ($55.6\%$). Analyzing and understanding information for incidents with potential harm to the organization is crucial for its ability to correctly focus its cyber security strategy and budget. We observed various views in our primary studies on how to analyze cyber threat situations. These views help analysts and decision makers to identify diverse aspects of the threats, including relationships between threats and assets [P4, P5], the status and progression of a threat [P2], and the evolution of threats [P7]. For example, researchers in [P5] provide views to the users to analyze the attack scenario in the form of an attack chain generated through attack scenario analysis of high-level threat alerts. These views allow the user to analyze how an attack occurs in the attack chain, identify any anomalies, and predict the next attack phase.

\begin{table}[]
\caption{\label{tab:datasources} Data source types}
\begin{tabular}{p{0.13\textwidth} p{0.24\textwidth} p{0.04\textwidth}}
\hline
\toprule
 Data source types &  Papers  & Count \\
\midrule

Security tools &  [P2, P5, P6, P7, P8, P12, P13, P16, P18, P19, P21, P26, P28, P31, P32, P39, P41, P52, P53]  &  19 \\ 
  Asset identification and management systems &  [P1, P5, P7, P20, P21, P22, P23, P24, P28, P29, P30, P31, P33, P34, P36, P37, P38, P39, P40, P42, P43, P45, P46, P54] & 24  \\ 
 External data source  & [P1, P6, P8, P9, P11, P16, P19, P20, P21, P23, P28, P30, P31, P41] &  14 \\
 Human input and organizational information  &  [P2, P3, P7, P8, P10, P12, P13, P17, P18, P20, P21, P22, P28, P31, P33, P36, P37, P51] & 18 \\
   Logs   & [P4, P10, P14, P16, P20, P21, P23, P25, P31, P33, P34, P35, P36, P37, P38, P40, P42, P44, P45, P47, P48, P49, P53] & 23 \\
     Network traces   & [P1, P4, P14, P15, P16, P18, P19, P20, P21, P22, P23, P24, P27, P29, P30, P31, P33, P50, P53]  &  19 \\
     
 \bottomrule
 
\end{tabular}

\end{table}
\subsubsection{What data sources are used} \label{sec:datasources}

This  section  presents  the  different data sources used for CSA visualizations in our selected set of primary papers.   The  data  extracted  for  this  section corresponds to item D7 in Table \ref{tab:dataextraction}. Our thematic analysis resulted in the identification of six types of data sources, as shown in Table \ref{tab:datasources}. We observed that multiple data sources are often used in the selected set of primary papers to generate CSA Visualizations. As a result, one paper can appear under two data sources in  Table~\ref{tab:datasources}.

 \textbf{Security tools:}
 We found 35.2\% of the primary papers in our SLR utilized information obtained from security tools. It includes  Security Information and Event Management (SIEM) [P2, P5, P8],  risk analysis  tools [P8], and output from various analysis tools [P16, P26].  For example, in [P5], researchers use high-level alerts generated by the correlation rule set defined in SIEM to represent nodes in the visualized attack scenarios. Researchers in [P8] use outputs of various risk analysis tools and  incident response trackers  for the proposed Cyber Common Operating Picture.

\textbf{Asset identification and management systems:}
One of the key aspects of cyber security is to systematically discover and select all relevant information assets that the organization holds; then,  potential security risks or gaps that affect them can be identified. 44.4\% of  primary papers utilize data from  asset identification and management systems. 
For example, the  Cyber COP system architecture  proposed in [P5] includes an asset database created using information gathered through \textit{asset identification and management systems}. In their architecture, various mechanisms such as a Simple Network Management Protocol (SNMP) and local agents  are used to gather asset information.
    
\textbf{External data sources:}
With the ever-increasing and complex cyber security incidents, organizations now need to move beyond data internal to the organization to make swift and effective cyber security decisions. Therefore, integrating external knowledge and data components is becoming an essential component for CSA. However, only 25.9\% of the primary papers use external data sources in their visualizations.  These data sources include common attack pattern enumeration, external domain sinkholing, GIS (Geographic Information System) maps, Malware sharing platforms, National vulnerability databases, and passive DNS systems. For example, in [P16], domain sinkholing strategies and a well-defined list of command and control server domains are adopted as the external data sources to identify networks with machines participating in botnet activities.

\textbf{Human input and organizational information:} 
Cyber security depends on various human inputs and organizational information. We observed that 33.3\% of papers use different human input and organizational information forms in their proposed CSA visualizations. 
The human input consists of   user-reported security incidents, expert knowledge, and security-related configuration parameters such as patching compliance ratings. Furthermore, organizational information includes mission dependencies and business processes. Researchers in [P17] use expert knowledge about risk profiles stored in text file format as input to their expert system to facilitate an institutional risk profile definition for CSA.

\textbf{Logs:} 
A log is a record of previous activities of a system, and the organization can use them to take corrective and preventive measures. For example, in a cyber incident case, logs can be used to identify what assets have been compromised and their severity. It is observed that 42.6\% of papers use logs as a data source for CSA visualizations. It includes database logs, firewall logs, IDS logs,  network logs, and web proxy logs.
    
\textbf{Network traces:} We found 19 (35.2\%) papers using network traces. This category consists of raw data and network data collected from tools such as Wireshark and Splunk [P1]. Furthermore, Network traffic [P19, P20, P21, P22, P23, P24, P29, P30, P31] and TCP/IP packet traces [P27] fall under this category. For example, in [P4, P16], the authors analyze data from network flow in addition to firewall logs and web proxy logs. In this context, a network flow represents an aggregation of packets exchanged by a pair of systems.

\subsubsection{How the cyber information is visualized}\label{visualizations}

\begin{table*}[t!]\caption{\label{tab:visualizationtechniques} Visualization techniques}
\begin{tabularx}{\textwidth}{lp{0.65\linewidth}p{0.65\linewidth}p{0.1\linewidth}}
\toprule
Visualization Techniques & References & Count \\
\midrule
Iconic displays & [P1, P2, P3, P4, P5, P6, P7, P8, P10, P11, P12, P13, P14, P16, P17, P18, P19, P20, P21, P22, P23, P24, P25, P26, P27, P29, P30, P31, P32, P33, P34, P36, P37, P39, P40, P41, P44, P45, P46, P47, P49, P50, P51, P52, P53, P54] & 46\\
Geometrically transformed displays & [P1, P2, P3, P4, P5, P7, P8, P9, P10, P11, P12, P13, P14, P15, P18, P19, P21, P22, P23, P24, P25, P28, P30, P31, P32, P33, P38, P39, P40, P43, P45, P46, P47, P49, P50, P51, P53, P54] & 38\\
Standard 2D/3D displays & [P2, P4, P9, P13, P14, P16, P17, P19, P21, P23, P25, P26, P29, P30, P34, P36, P37, P38, P40, P41, P42, P44, P47, P48, P49, P50, P53] & 27\\
Tables/text summaries & [P2, P4, P5, P10, P12, P13, P16, P22, P26, P28, P30, P31, P32, P34, P35, P36, P37, P38, P39, P42, P45, P48, P49, P51, P54] & 25\\
Geographical displays & [P1, P4, P5, P8, P12, P13, P16, P19, P23, P26, P30, P31, P34, P38, P39, P41, P42, P52, P54] & 19\\
Stacked displays & [P6, P18, P20, P21, P28, P32, P33, P36, P40, P43, P44, P49, P50] & 13\\
Single value displays & [P8, P13, P34, P37, P38, P39, P40, P41, P42, P43, P48, P51, P54] & 13\\
Dense displays & [P4, P8, P13, P19, P20, P27, P44, P47, P49, P50] & 10\\
Immersive environments & [P8, P11, P12, P39, P40] & 5\\
\bottomrule

\end{tabularx}
\end{table*}

\Figure(topskip=0pt, botskip=0pt, midskip=0pt) [width=0.9\linewidth] {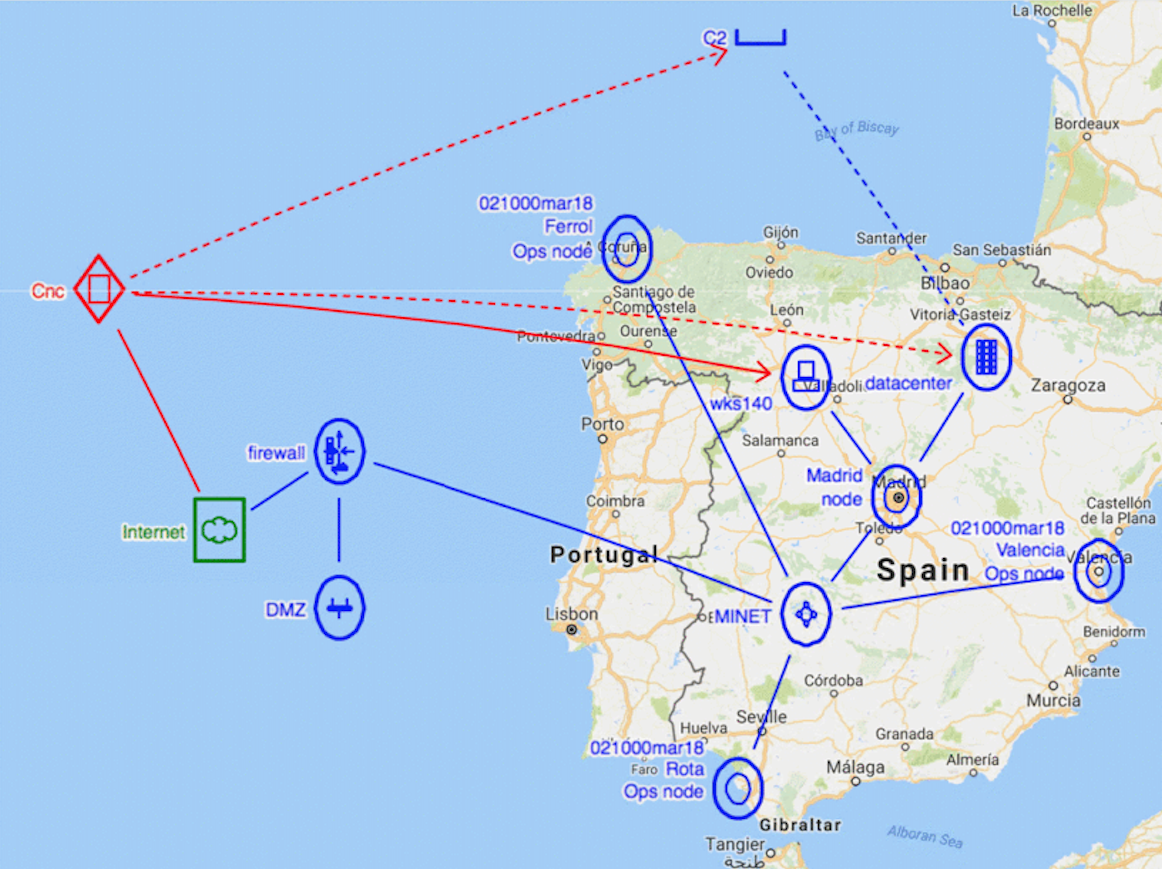}
{The visualization for Improved Situational Awareness (VISA) demonstrator employed in [P8] provides a common  operational picture to military staff. The visualization employs traditional military symbols.\label{pid30_VISA}}

\Figure(topskip=0pt, botskip=0pt, midskip=0pt) [width=0.9\linewidth] {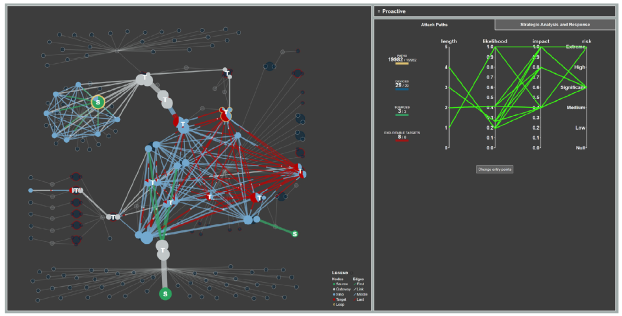}
{The visualization proposed in [P2] provides two views for the proactive environment (i.e., two visualizations): i) view  on the left shows the network topology; and ii) view on the right shows a summary of information related to the attack graph and parallel coordinates visualization to support its analysis.\label{PID7_proactiveenvironment}}

\Figure(topskip=0pt, botskip=0pt, midskip=0pt) [width=0.9\linewidth] {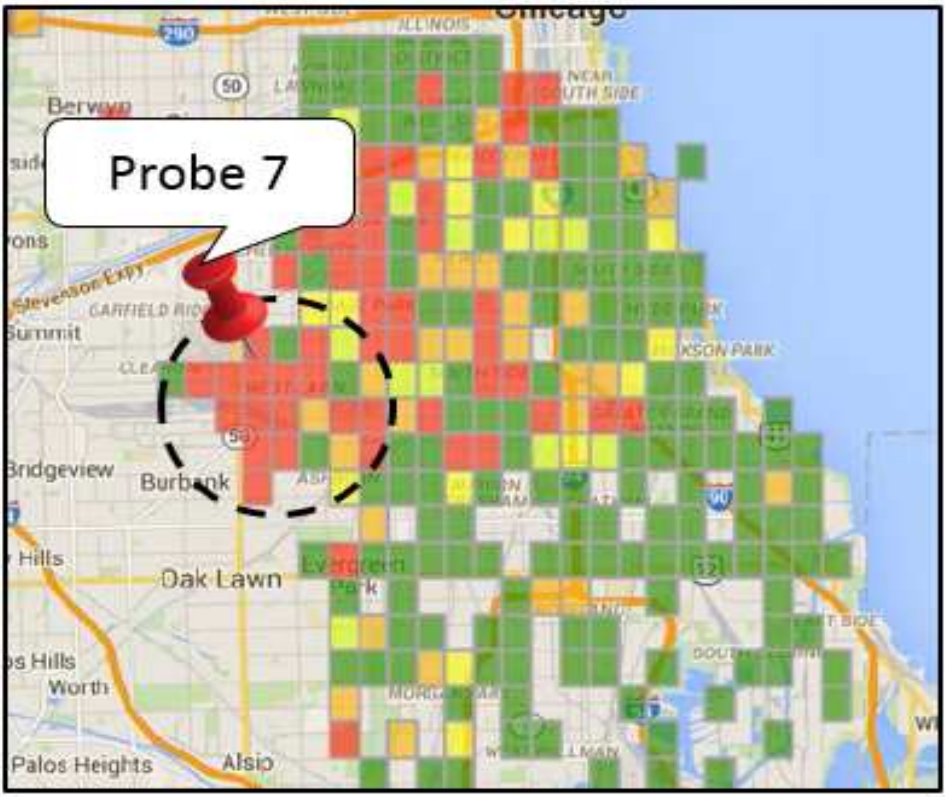}
{Visualization proposed in [P19] uses heatmaps to determine the general location of a Field Area Network (FAN) from where the anomalous traffic is emanating. Here dashed circles indicate possible problematic areas.\label{pid90_heatmap}}

\Figure(topskip=0pt, botskip=0pt, midskip=0pt) [width=0.9\linewidth] {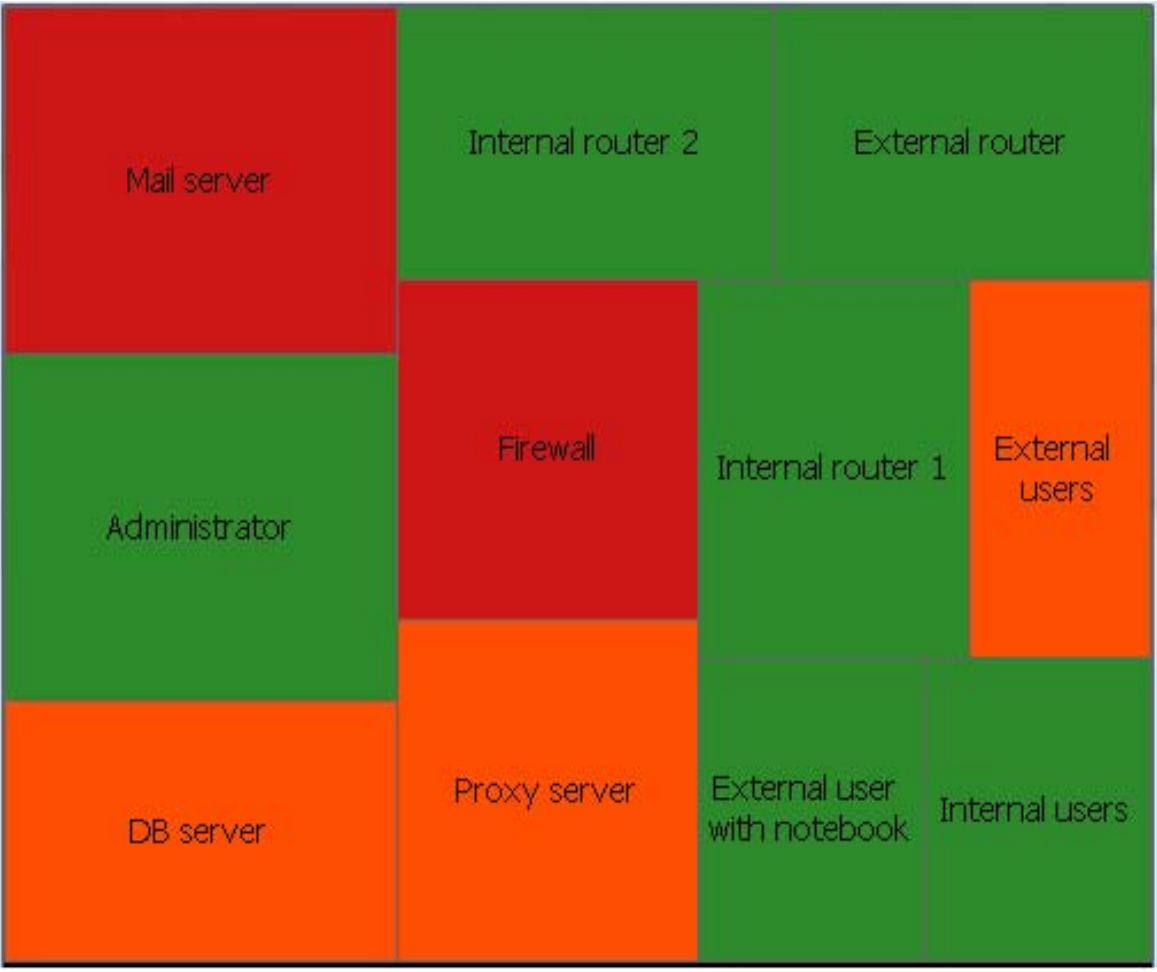}
{Visualization proposed in [P21] is a treemap.\label{pid107_treemap}}

\Figure(topskip=0pt, botskip=0pt, midskip=0pt) [width=0.9\linewidth] {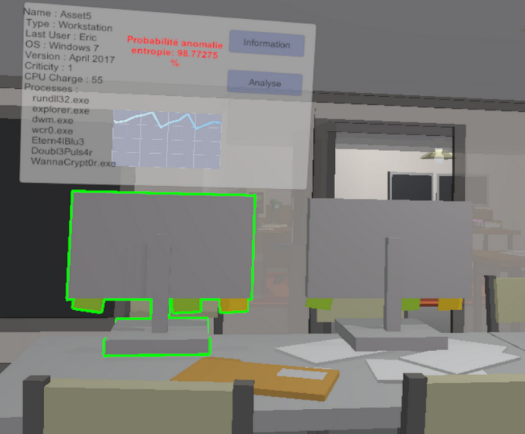}
{Visualization proposed in [P39] is in an \textit{immersive environment}.\label{ppid27_immersive}}

\Figure(topskip=0pt, botskip=0pt, midskip=0pt) [width=0.9\linewidth] {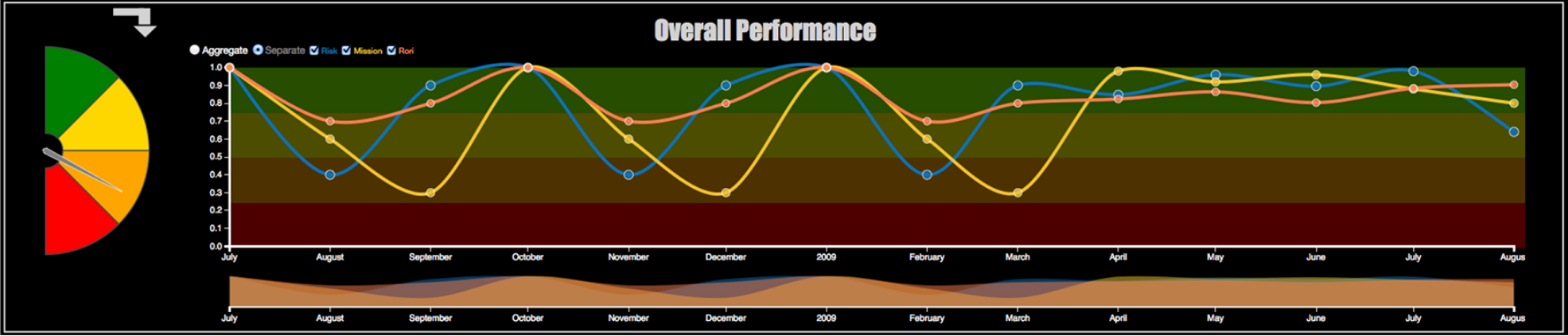}
{Visualization proposed in [P13] proposes a tachometer view to facilitate financial security managers get an overall view of the system performance. Furthermore, the view also provides trends and patterns of various indicators.\label{pid52_gaugue}}

\Figure(topskip=0pt, botskip=0pt, midskip=0pt) [width=0.9\linewidth] {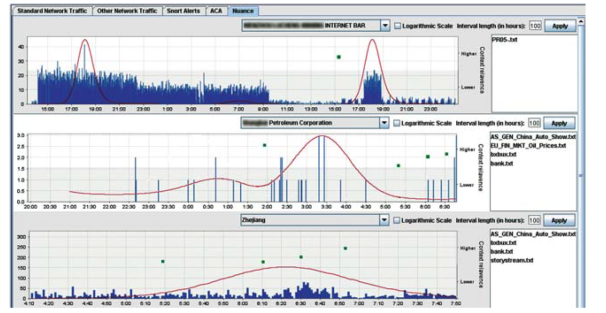}
{Visualization proposed in [P30] uses standard 2D bar and line charts.\label{pid146_2D}}

\Figure(topskip=0pt, botskip=0pt, midskip=0pt) [width=0.9\linewidth] {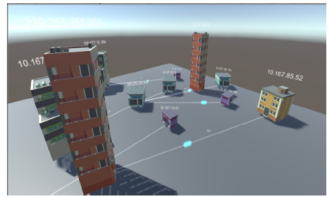}
{Metaphoric visualization proposed in [P1] uses the buildings and cars as metaphors to show the network activity to non-expert users. It uses the relative positioning to show the IP addresses.\label{pid3_metaph}}

CSA visualizations employ various visualization techniques. Furthermore, diverse \textit{tasks/interactions} are linked with CSA visualizations to improve user experience. In this section, we report how the CSA information described before is visualized considering the visualization techniques and related \textit{tasks/interaction} techniques (i.e., the data extracted for this section corresponds to items D8 and D9 in Table 3).

\paragraph{Visualization techniques}\label{ss:visualization techniques}

 Table~\ref{tab:visualizationtechniques} presents how the  \textit{visualization techniques} are distributed over the selected studies.
 We describe these categories below.
 
\textbf{Iconic displays}: \textit{Iconic displays} are the most common class of visualization techniques reported in the studies considered in this  SLR (85.2\%).  In \textit{iconic displays}, the attributes of multidimensional data items are mapped onto the features of an icon for the representation. Some of the common \textit{iconic displays} reported in our primary studies include color icons [P2, P4, P13, P14, P16, P18] and shape icons [P8, P12, P21]. In addition, color icons often highlight the importance and significance of the reported values [P13, P20, P22, P25, P26]. 
Furthermore, some primary studies associate icon sizes with numerical attributes. For example, in [P5], the node's size is used to represent its importance to the respective mission. Several studies use special icons familiar to the user in their visualizations.  For example, the work reported in [P1] uses shape icons that are familiar to the user, such as laptops, cables, buildings, and roads, to visualize the cyber status (see Figure~\ref{pid3_metaph}). 

 \textbf{Geometrically transformed displays}:
 Often cyber security data consists of more than three attributes and, therefore, they do not allow a simple visualization as 2D or 3D plots described previously. This category includes visualizations that use interesting transformations of multidimensional data sets. We found that 70.4\% of the primary studies in our SLR  use \textit{geometrically transformed displays} in their visualizations.
    Common examples are the  node-link diagrams [P2, P3, P5, P12, P31] and
	parallel coordinate plots [P2].  Parallel coordinates plot each multidimensional data item as a polygonal line that intersects the horizontal dimension axes at a position corresponding to the data value for the corresponding dimension (see Figure~\ref{PID7_proactiveenvironment}). We also observed other visualizations with interesting transformations of multidimensional data. For example, in an area corruption chart proposed in [P13], each compromised device produces a hole in the area representing the supported sub-process. The hole is proportional to the value of its operational impact score. Furthermore, the Mission-Attacker-Controls triangle (MAC) proposed in [P8] is a 3D triangular plot used to show the relative forces of the mission, the attacker's interest in the asset, and the security controls.

 \textbf{Standard 2D/3D displays}: A large number of the primary studies selected for this SLR use \textit{standard 2D/3D displays} (50.0\%).  This includes visualization techniques like x-y plots (e.g., scatter plots [P14, P21, P25], bar charts [P4, P26, P29, P30], pie charts [P29]), and line charts [P9, P17, P30]).
    For example, the works reported in [P26, P30] use bar charts to illustrate the distribution of the standardized incidence rate and the per-minute observed traffic levels, respectively.
    In [P25], scatter plots show similar alerts grouped where an alert is represented as one dot in the visual space.

\textbf{Tables/text summaries}: We identified tables and text summaries as a popular form of presenting cyber information in the selected papers (46.3\%). 
 \textbf{Geographical displays}:   In our selected set of primary studies, 35.2\%  of the studies use \textit{geographical displays} to visualize geographical information.  
 Maps are often employed to present the geographical distribution of information related to assets [P5, P8, P31], risks [P16, P19, P23], and threats [P5, P8].  For example, researchers in [P13] use maps to illustrate how the network nodes are geographically distributed. The work reported in [P16] employs maps to illustrate how the attacker capacity is distributed worldwide and provides the user with a closer look at the organizations infected by malware.  In [P20],  maps are used to display cities with extremely high or low malicious activities.
 
\textbf{Stacked displays}: \textit{Stacked displays} are representations of hierarchical data and hierarchical layouts for multidimensional data. However, only a limited number of studies (24.1\%) use this display in their visualizations. Treemaps are examples of a hierarchical data representation found in two papers [P20, P21], displaying hierarchical data as a set of nested rectangles. 
For example, in the treemap visualization proposed in [P21] (see Figure\ref{pid107_treemap}), the business value of the host defines the rectangle size, and the calculated host security level defines the color.  Another example of \textit{stacked displays} can be found in work reported in [P6], where the risk levels are visualized using a risk tree visualization. In [P28], a tree view represents the entire attack graph in the form of a directory hierarchy.  In work reported in [P33], a hierarchy of layers presents how types of operational missions, mission-critical tasks, and types of assets are connected together. 
	
\textbf{Single value displays}: 
\textit{Single value displays} show an interesting representation of a single or instantaneous value that is meaningful on its own. We observed this visualization in  24.1\% of the selected set of papers. Gauge representations are a common form of \textit{single value displays} [P8, P13]. Researchers in [P13] (see Figure~\ref{pid52_gaugue}) use gauge representations to provide a glance view of several performance indicators to the financial security manager. 

\textbf{Dense pixel displays}:  Each data point in \textit{dense pixel displays}  is mapped to a colored pixel so that they can be  grouped into adjacent areas that represent individual data dimensions. However, only a few  studies (18.5\%)  use this type of display. A heatmap is an example of dense pixel displays employed in studies reported in this SLR [P4, P8, P13, P19, P20, P27]. A heatmap is a two-dimensional representation of data in which values (i.e., the magnitude of phenomena) are represented by different colors (see Figure~\ref{pid90_heatmap}).

 \textbf{Immersive environment}: 
\textit{Immersive environments} allow users to immerse themselves in the artificially-created virtual environments through a collection of computer hardware and software so that users could perceive themselves to be included in and interact in real-time with the environment and its contents. However, only a limited set of studies  (9.3\%) employ  \textit{immersive environments}  in their  visualizations [P8, P11, P12, P39, P40]. 
In [P8, P11, P12], virtual reality head-mounted displays are used to create an illusion for the user of immersion in virtual cyberspace. In [P39], a Collaborative Virtual Environment is deployed for the 3D Cyber COP model to help cyber analysts mediate analysis activities. The \textit{immersive environment} is shown in Figure~\ref{ppid27_immersive}. 
Through the use of these environments, users have the impression that they are inside an environment rather than viewing it from the outside.

Multiple visualization techniques are often utilized together in a single visualization.
We observed that apart from  \textit{standard 2D/3D displays} and \textit{tables/text summaries}, other visualization techniques are combined with \textit{iconic displays} in more than 50\% of the visualization instances of our selected set of papers. \textit{Iconic displays} place the information en-richer role in most of these visualizations. For example, color or shape icons are often  used with  \textit{geometrically transformed displays},  \textit{geographical  displays}, and  \textit{stacked displays} to emphasize the status, severity, or impact of a particular phenomenon [P1, P5, P16, P22, P23, P31, P33].

Besides \textit{iconic displays}, \textit{geometrically transformed displays} are often combined with other visualization techniques.
For example, the work reported in  [P1, P11] combined a node-link diagram (an example of a \textit{geometrically transformed display}) with \textit{immersive environments}.
In these examples, the users can immerse in the environment through the virtual reality headsets to investigate the properties of the node-link diagrams.  It is also interesting to note that \textit{single value displays} instances are used in combination with \textit{standard 2D/3D displays} more than 50\% of the time. When source literature is referred to, it is clear that \textit{standard 2D/3D displays} are used to provide additional information to interpret the metrics visualized through the \textit{single value displays}. For example, in  Figure~\ref{pid52_gaugue},  a \textit{standard 2D display} shows the trends and patterns of the associated metrics while the tachometer shows the overall system performance. 

\begin{figure} 
\centering
\includegraphics[width=0.48\textwidth]{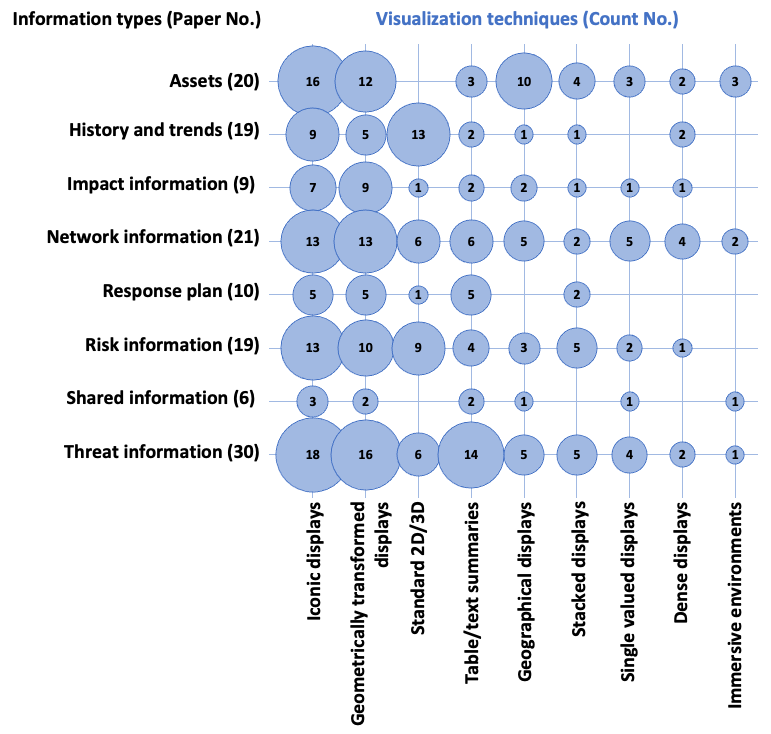}
\caption{Visualization techniques vs information types}
\label{vistech_infotypes}
\end{figure}

We also compared the cyber security  information types discussed in Section~\ref{sec:whatInfo} with the utilized visualization techniques.
According to  Figure~\ref{vistech_infotypes}, it is evident that all the information types often employ  \textit{iconic displays}  as a  visualization technique. Furthermore, \textit{geometrically transformed displays}  are often  employed in visualizations that present information related to networks [P5, P11, P14, P15, P24],  assets [P1, P5, P8, P13, P22], risks [P14, P18, P21, P23, P32],  threats [P13, P19, P32],  and impact [P8, P9, P13, P21, P24].  When presenting asset information,  \textit{geographical displays} are often employed as a visualization technique [P5, P8, P13, P31]. 
\textit{Standard 2D/3D displays} are often employed to visualize information related to history and trends [P25, P27] and risk evaluation [P16, P17]. \textit{Tables/text summaries}  are mainly used to convey information related to threats [P5, P22, P31, P32] and response plans [P2, P16, P28]. Furthermore,  \textit{immersive environments}  commonly visualize network information [P11, P12].

\begin{table}[]\caption{\label{tab:taskinterractions} Task/interaction techniques}
\begin{tabularx}{0.48\textwidth}{lp{0.52\linewidth}p{0.52\linewidth}p{0.1\linewidth}}
\toprule
Tasks/interactions & References & Count \\
\midrule
Zooming & [P2, P7, P8, P13, P20, P22, P23, P28, P32, P40, P46, P47, P48, P53] & 14\\
Filtering & [P2, P3, P4, P7, P13, P16, P25, P26, P28, P35, P38, P40, P41, P44, P45, P46, P47, P48, P49, P53] & 20\\
Details on demand & [P2, P12, P13, P16, P19, P21, P25, P28, P30, P40, P41, P42, P43, P44, P45, P46, P47, P49, P50, P51, P53] & 21\\
History & [P2, P7, P16, P44, P49, P50, P53, P54] & 8\\
Relate & [P2, P4, P5, P7, P40, P53] & 6\\
Extract/share & [P4, P16, P28, P31] & 4\\
Move/rotate & [P8, P11, P12, P51] & 4\\
Linking/brushing & [P50] & 1\\
Customization & [P6, P8, P28, P37, P38] & 5\\
\bottomrule
\end{tabularx}
\end{table}

\paragraph{Tasks/interactions} \label{ss:interaction techniques}
Interaction techniques allow users to interact with the visualizations and facilitate effective data exploration directly. Table~\ref{tab:taskinterractions} illustrates the tasks/interactions type distribution over the selected set of papers.
 We also point out that there are 16 papers (29.6\%) that we did not classify into \textit{tasks/interactions} topics.  A similar observation was made previously in \cite{guimaraes2015survey}. Similar to \cite{guimaraes2015survey}, we are unsure whether the authors do not highlight these features or the proposed visualization does not provide such features. The \textit{tasks/interactions} classification used in this paper is described below.

\textbf{Zooming}:  \textit{Zooming} helps to present data in a highly compressed form to provide an overview of the data while at the same time allowing a flexible display of the data at different resolutions based on the user's needs.  As evident from Table~\ref{tab:taskinterractions}, \textit{zooming} functionality is indicated in 25.9\% of the publications. This functionality allows users to zoom in on items of interest to them. For example, in [P8], an operational picture of the situation is initially presented at a higher level of abstraction, and when the users zoom in, abstract nodes are replaced by their detailed representations.

\textbf{Filtering}: \textit{Filtering} allows users to interactively partition the data set into segments and focus on interesting subsets. In our selected set of primary studies, 37.0\% of the papers use the \textit{filtering} functionality. For example, the parallel coordinates visualization proposed in [P2] allows users to filter a set of attack paths by brushing on one or more axes.  The \textit{filtering} allows a complex set of attack paths to be quickly reduced only to paths that respect certain conditions defined by the user. In [P25], the proposed visualization allows the user to filter cyber events based on several criteria.

\textbf{Details on demand}: \textit{Details on demand} functionality allows users to select an item or group and get details when needed.  This functionality is mentioned in   38.9\% of the primary papers.  \textit{Details on demand} are often provided by clicking or double-clicking options [P13, P16, P21, P28, P30] and tool-tips [P2, P25].

 \textbf{History}: \textit{History} allows support users to keep and view the history step by step through different options such as undo and replay. However, as evident from Table~\ref{tab:taskinterractions}, this functionality is only mentioned in eight publications (14.8\%). 
 
 \textbf{Relate}: This enables viewers to view relationships among items.  Through the \textit{relate} interaction, users can click on one item and see its relationships to other items.  Only six (11.1\%) publications included this user interaction.

\textbf{Extract/share}: This allows users to share item(s) that they desire with others or extract item(s) that they desire for later use. After extracting, the users could save the data to a file in a format that would facilitate other uses such as sharing, printing, and graphing [P4, P16]. We found only four papers (7.4\%) in this category.

 \textbf{Move/rotate} Moving and rotating the visualization [P8, P11, P12]. Moving and rotating are related to 3D displays and immersive environments. However, we only observed four papers (7.4\%) that discussed this user interaction.

\textbf{Linking and brushing}: \textit{Linking and brushing} allows  interactive changes made in one visualization to be reflected automatically in other visualizations. However, we only found one paper (1.9\%) that mentioned this ability. In [P50], analysts are allowed to make good use of both heatmaps and line charts to overcome their weaknesses by implementing \textit{linking and brushing} interactions.

\textbf{Customization}: Editing the visualization (edit mode). We observed this user interaction in only 5 (9.3\%) publications. In both [P5] and [P28], users can customize the visualizations by changing the layout. In [P6], users are given a series of visualization techniques to choose from. Here the user can pick the visualization technique that best suits them to visualize the information at hand.

\begin{table}[] \caption{\label{tab:CSALevelref} Cyber Situational Awareness Level}
\begin{tabularx}{0.98\linewidth}{p{0.2\linewidth}p{0.55\linewidth}p{0.1\linewidth}}
\toprule
Situational Awareness Level & References & Count\\
\midrule
Perception & [P1, P2, P4, P5, P6, P8, P9, P10, P11, P12, P13, P14, P15, P16, P17, P18, P19, P20, P21, P22, P23, P24, P25, P26, P27, P28, P29, P30, P31, P33, P34, P35, P36, P37, P38, P39, P40, P41, P42, P43, P44, P45, P46, P47, P48, P49, P50, P51, P52, P53] & 50\\
Comprehension & [P2, P3, P4, P5, P7, P8, P13, P16, P19, P20, P21, P22, P23, P24, P26, P28, P30, P31, P32, P34, P36, P37, P38, P39, P40, P43, P44, P46, P54] & 29\\
Projection & [P2, P5, P8, P16, P21, P24, P28, P39, P43, P54] & 10\\
\bottomrule
\end{tabularx}
\end{table}

\subsection{RQ3:  What  level of CSA is facilitated by the visualizations?} \label{CSAsupport}

\vspace{1em}
\begin{mdframed}
     \begin{itemize}
  \item  Most studies (92.6\%) facilitate the  \textit{perception}  level, and several studies (53.7\%) facilitate up to  \textit{comprehension} level. 
   \item Only a limited number of studies  (18.5\%) provide visualizations to achieve up to \textit{projection} level.
\end{itemize}
\end{mdframed}

As described in  Section~\ref{background:CSA}, the Endsley~\cite{endsley1995toward} model   provides three ascending levels of SA, namely \textit{perception}, \textit{comprehension}, and \textit{projection}, which may or may not be linear.  In this section, we analyze what levels of CSA,  described in Section~\ref{background:CSA},  can be achieved through the proposed visualizations.  It is also important to highlight that some publications included in this SLR provide multiple visualizations that may facilitate achieving multiple SA levels. 
In the case where a single visualization can be used to achieve multiple levels of CSA, we assigned the corresponding highest level of CSA for that particular visualization. 
Table~\ref{tab:CSALevelref} illustrates the levels of CSA supported by the publications selected in this SLR and presents the distribution of papers across the three CSA levels. As one publication could provide multiple visualizations, in Table~\ref{tab:CSALevelref}, a single publication can be reflected in multiple levels of SA.   

\Figure(topskip=0pt, botskip=0pt, midskip=0pt) [width=0.9\linewidth] {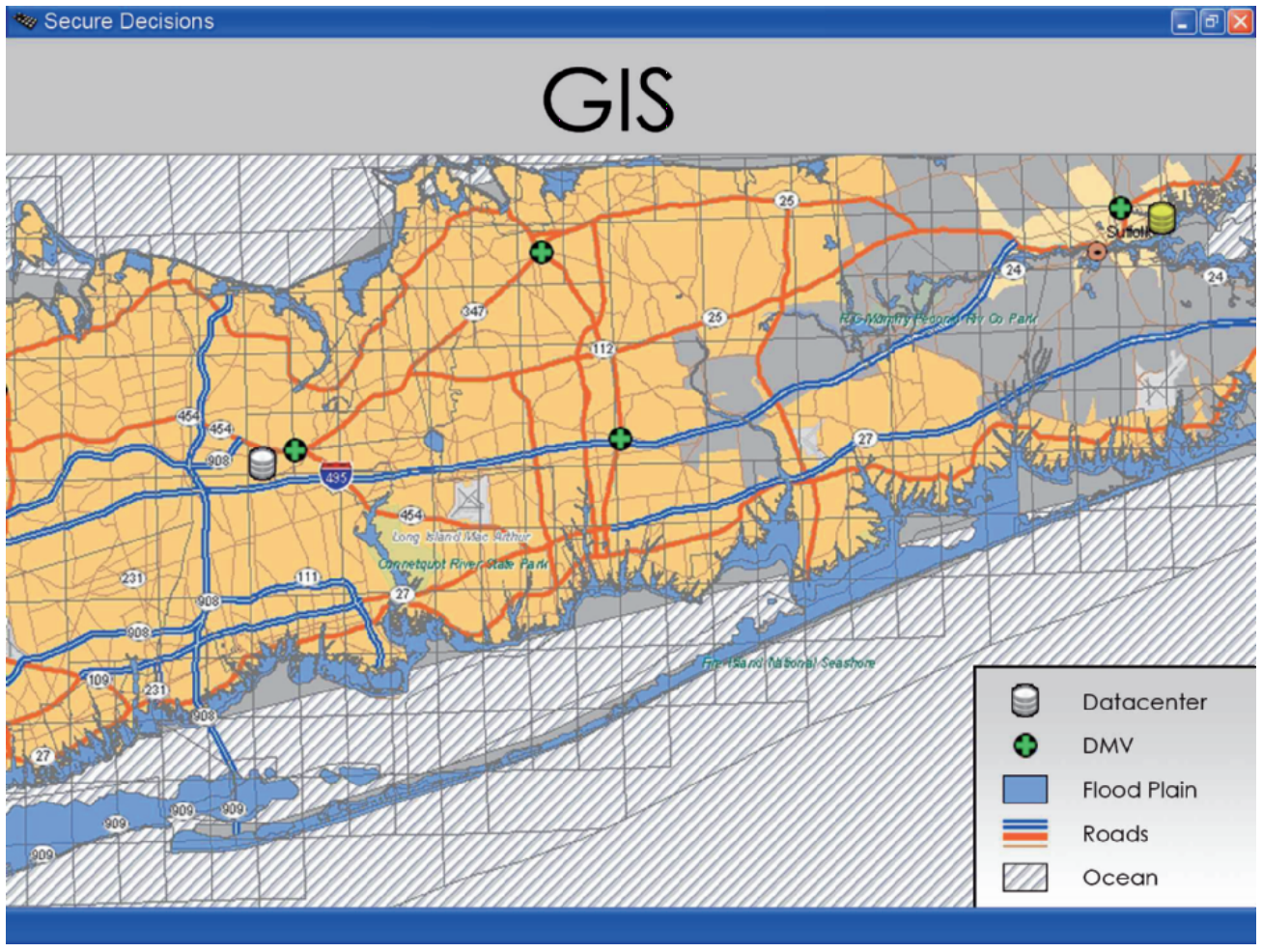}
{Visualization proposed in [P31]  provides the user an overview of the geographical distribution of critical infrastructure.\label{pid52_areacorruption}}

\Figure(topskip=0pt, botskip=0pt, midskip=0pt) [width=0.9\linewidth] {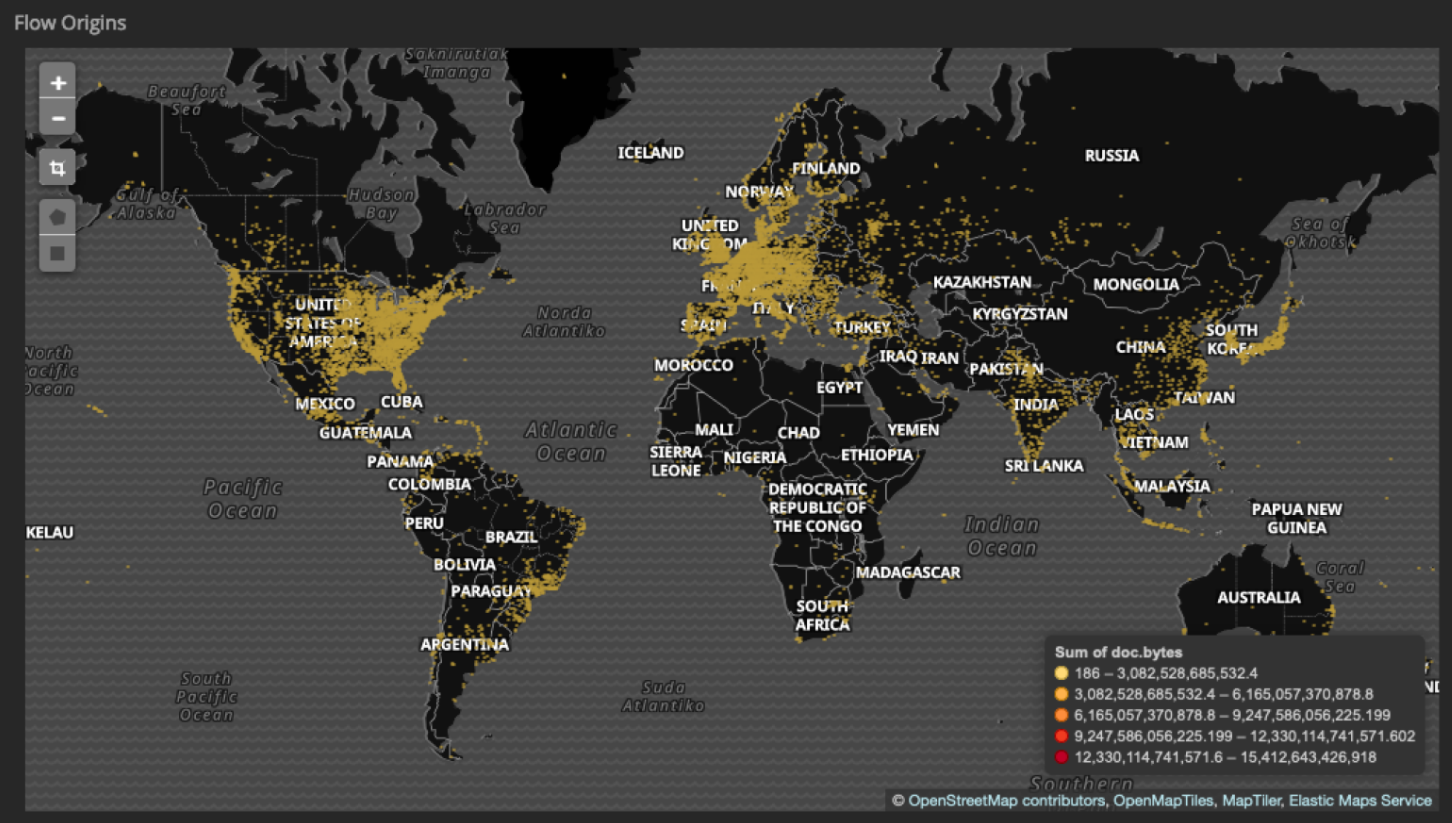}
{Visualization proposed in [P38] shows information about the location of source IP addresses.\label{perceptionvisualizations}}

\paragraph{Perception}
Visualizations that provide users with an overview of the status, attributes, and dynamics of the cyber environment have been linked to the \textit{perception} level (see Figure~\ref{perceptionvisualizations}). These visualizations often allow the user to answer the question ``What is happening in the cyber environment?''. Most studies in this category provide visualizations that present a high-level overview of the cyber assets [P5, P20, P31, P33],  network topology [P2, P29],  cyber threats [P4, P5, P25, P26], and cyber risks [P2, P14, P16, P28].  For example, Cho et al.  [P5] propose a geographical perspective view that allows the user to identify the status of cyber assets and threats. Carvalho et al. [P16] provide an indication of the attacker's capacity by visualizing the distribution of bots over a world map. Kopylec et al. [P31] provide the user with an overview of the geographical distribution of critical infrastructure using maps.  
Angelini et al. [P2] provide a visualization to allow users to obtain an overview of the system's network topology and risk status. For this, they have superimposed an attack graph over the network topology.  Using the attack graph, the user can get an overview of the risk posture of the organization (see Figure~\ref{PID7_proactiveenvironment}).   Yu et al.  [P26] use a world map to show cities with the highest Standardized Incidence Rate (SIR). The SIR metric can be used to  identify cities with higher infection levels and is defined as the "number of malicious IP addresses for every 100,000 actual machines that could be infected in a city". 

\Figure(topskip=0pt, botskip=0pt, midskip=0pt) [width=0.9\linewidth] {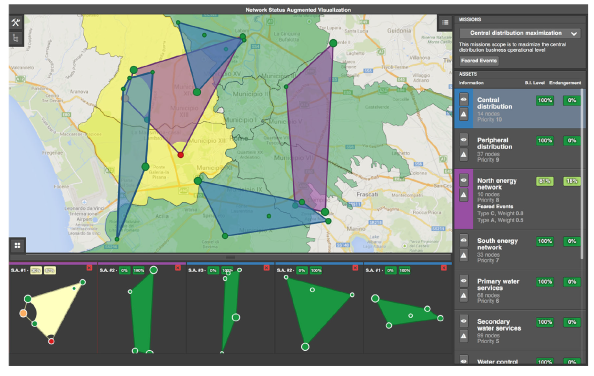}
{Visualization proposed in [P13] shows the impact of compromised nodes through the concept of area corruption.\label{pid52_areacorruption}}

\Figure(topskip=0pt, botskip=0pt, midskip=0pt) [width=0.9\linewidth] {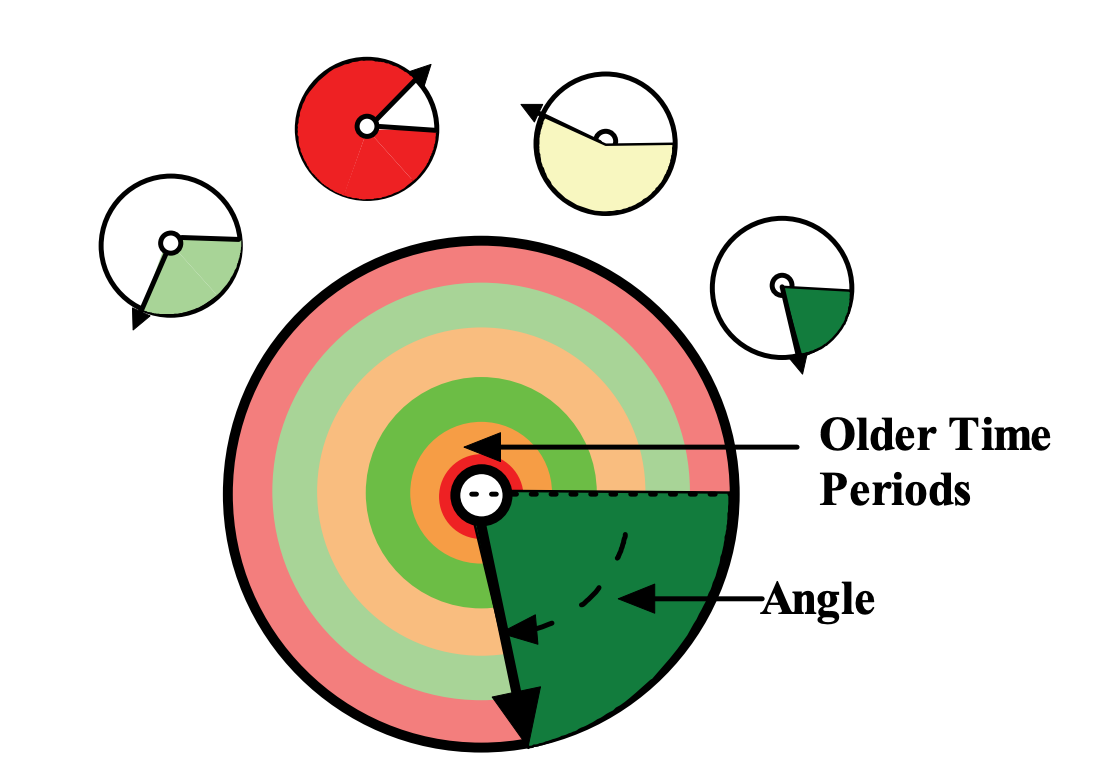}
{Visualization proposed in [P24] to allow decision makers to interpret the cyber situation. History information is included in the visualization. Small dials provide more information on the individual components of the system. The dial is reinforced to facilitate rapid interpretation.\label{pid120_gauge}}

\paragraph{Comprehension}
  \textit{Comprehension} allows users to move from simply being aware of the elements in the cyber environments to comprehending the situation. Therefore, visualizations that facilitate the users to understand the meaning of the elements in the cyber environment are linked to this category  (see Figure~\ref{pid52_areacorruption} and Figure~\ref{pid120_gauge}). These visualizations allow the user to answer questions like  ``Why it is happening?'' and ``What is the meaning?'' with respect to elements in the cyber environment.  Several studies in this category provided visualizations that provide the context of the elements in the cyber environment [P2, P4, P5, P24, P32]. %Providing the context allows the user to move from just being aware of a situation to comprehending its meaning.  
 For example, the work reported in [P4] provides an `event detail page' that provides the context of a selected cyber event. It includes horizon graphs of several flow fields and heatmaps of IP addresses that provide temporal context to the event. These visualizations prioritize showing trends and patterns since this is most important for context. Understanding the context of a specific event allows users to comprehend its meaning, and this can be considered a higher mental state than simply being aware that a cyber incident has occurred. Authors in [P24] propose a visualization by extending standard gauge visualizations (see Figure~\ref{pid120_gauge}). Their visualization includes a large dial and a set of smaller dials that show the system's overall status, network, or mission and how individual system components are being impacted. The information provided in the smaller dials  provide context to understand the information shown on the large dial. Furthermore, to provide more context into what is shown on the larger dial, history information has been added by providing rings within the dial where the outer ring shows the current value.  The work reported in  [P5] allows us to see how a specific attack has been taking place in terms of five attack phases of a proposed attack chain model. It allows users to closely investigate the attack progression and take actions if needed. 

 Some visualizations in the \textit{comprehension} category also specifically looked at providing information on the significance/consequence of cyber incidents to the user [P8, P13, P28].  Understanding the impact, significance, or consequence of the cyber incidents allows users to comprehend the situation and is a higher mental state than being just aware of the cyber incidents that have occurred. For example, the work reported in [P28] visually displays the effects that occur when a specific node or protection domain is affected.  This allows users to move from just being aware of the threat situation to understanding and comprehending the threats with respect to organizational goals. In [P13], the impact of a compromised device on its supported process is shown through the concept of area corruption. The idea is to have a hole in the area representing the supported sub-process for each compromised device. The  hole  is proportional to the value of its operational impact score. It allows users to understand the significance of the cyber incidents and their relationships to the supported process. 
 
\Figure(topskip=0pt, botskip=0pt, midskip=0pt) [width=0.9\linewidth] {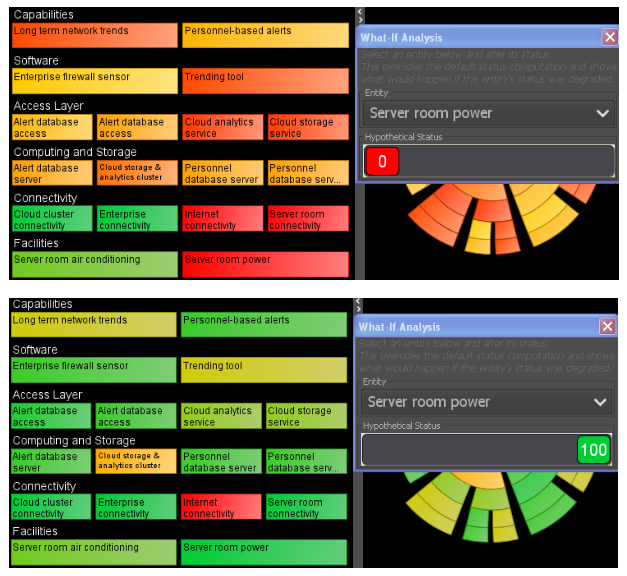}
{What-if analysis result proposed in [P43] provides a predictive analytics capability.\label{pid30_MAC}}

\Figure(topskip=0pt, botskip=0pt, midskip=0pt) [width=0.9\linewidth] {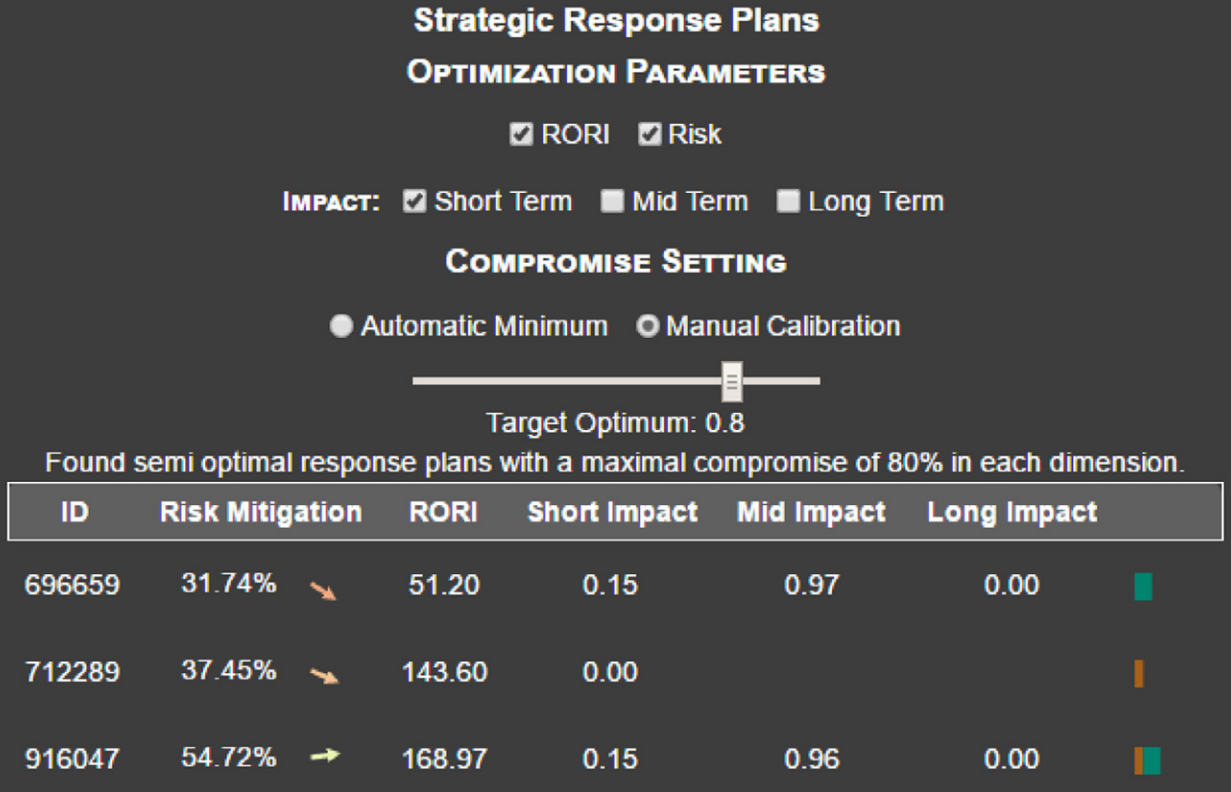}
{In [P2],  response plans are shown in a table and classified by their characteristics.\label{projectionvisualizations}}

\paragraph{Projection}
The visualizations that facilitate the \textit{projection} level allow users to predict the future cyber threat situation and possible future actions.  These visualizations allow the user to answer questions like  ``What will happen next?'' and ``What can I do?'' with respect to the cyber environment. A number of studies in this category have provided visualizations to illustrate the impact of possible future threats [P8, P21, P24] and possible plans to respond to the cyber security situation of the organization [P2, P21]  (see Figure~\ref{pid30_MAC} and Figure~\ref{projectionvisualizations}).     For example, the work reported in [P8, P28, P43] facilitates what-if analysis to assist the user in identifying possible future actions. Through the what-if analysis proposed in [P28], the user can specify a starting point for the attack (the presumed threat source), as well as an attack goal (critical network asset to protect). The results of what-if analysis allow the user to model the effects of software patches or other mitigation solutions on the system. In [P8], what-if analysis will allow the decision maker to analyze different action plans based on the importance given to the mission, the attacker's interest in the asset, and the security controls. Furthermore, the system also provides recommendations for optimal network defense. In [P2], response plans are shown to the user based on the current cyber situation. The proposed visualizations also allow users to understand how each response plan could reduce the risk on the network devices. Kotenko and Novikova [P21] visualize the \textit{Return-on-Security-Investment} index for each countermeasure that characterizes possible damages due to the security incident and the cost of security incidents.

 \begin{figure} 
 \centering
 \includegraphics[width=0.5\textwidth]{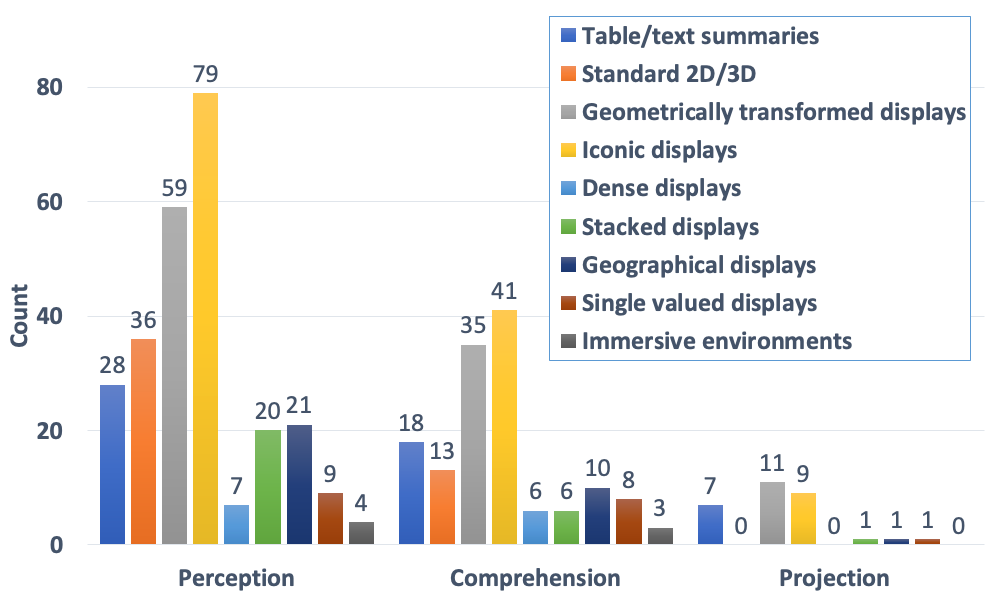}
 \caption {Visualization techniques for achieving different levels of SA}
 \label{vistech_sa}
 \end{figure}

We also analyzed the distribution of visualization techniques (described in Section~\ref{ss:visualization techniques}) with respect to the three levels of CSA  (see Figure~\ref{vistech_sa}).  According to Figure~\ref{vistech_sa}, \textit{iconic displays} are the most commonly used visualization technique at the \textit{perception} and \textit{comprehension} level.  At the \textit{projection} level, the most popular  visualization technique is \textit{geometrically transformed displays}.  Furthermore, it is interesting to note that popularity for \textit{standard 2D/3D displays} and \textit{geographical displays} gradually reduces over the CSA levels where there are no \textit{standard 2D/3D displays} and \textit{geographical displays} at the \textit{projection}  level.

\subsection{RQ4: What is the maturity of the proposed visualizations that facilitate cyber situational awareness? } \label{maturity}

\vspace{1em}
\begin{mdframed}
     \begin{itemize}
     \item Most studies use demonstrations or toy examples in their evaluations.
  \item Unfortunately, most proposed CSA visualizations lack rigorous and industry-suitable evaluation.
\end{itemize}
\end{mdframed}

\begin{table}\caption{The evidence is available in the selected studies to adopt the proposed visualization techniques.} \label{tab:maturitylevels} 
\begin{tabular}{p{0.3\linewidth}p{0.45\linewidth}p{0.1\linewidth}}
\hline
\textbf{Maturity level}  & \textbf{References} & \textbf{Count}\\ \midrule
No evidence & [P5, P13, P18, P28, P37]  & 5\\ 
Demonstration or toy examples & [P6, P7, P8, P9, P11, P12, P14, P17, P19, P20, P21, P22, P23, P25, P26, P27, P29, P31, P33, P35, P43, P44, P47, P48, P49, P52, P53, P54]  & 28 \\ 
Expert opinions or observations & [P1, P2, P30, P36, P40, P45, P46, P50] & 8\\ 
Academic study & [P3, P39, P47] & 3 \\ 
Industrial case study (casual case study) & [P10, P15, P24, P32, P34, P41]   & 6\\ 
Industrial practice& [P4, P16, P38, P42]  & 4 \\ \hline
\end{tabular}  
\end{table}

To answer RQ4, we analyzed the data collected for D11  in Table~\ref{tab:dataextraction}. The importance of rigorous evaluation to assess the appropriateness of the proposed solutions has been emphasized by the software engineering research community \cite{zannier2006success}. As mentioned in Section~\ref{dataanalysis_taxanomies}, we used a six-level hierarchy proposed in \cite{alves2010requirements} for assessing the reported evidence.   The proposed six-level hierarchy is listed below: i) no evidence; ii) evidence obtained from demonstration or working out with toy examples; iii) evidence obtained from expert opinions or observations; iv) evidence obtained from academic studies (e.g., controlled lab experiments); v) industrial studies (e.g., casual case studies); and vi) evidence obtained from industrial practice. This hierarchy has been used in previous studies to evaluate the maturity of visualizations in other domains~\cite{shahin2014systematic}. In particular, `no evidence' and `demonstration or toy
examples' are at the weak end of the hierarchy, while `industrial practice' indicates that the method has already been approved and adopted by an organization which may indicate convincing proof that something works. 

Table~\ref{tab:maturitylevels} presents the distribution of the studies according to the six levels of evidence. 
From Table~\ref{tab:maturitylevels}, it is evident that five studies (9.3\%) do not have any evaluation evidence of the proposed visualizations. Most of the primary studies (51.9\%) selected in this SLR show their maturity through demonstrations or toy examples. Some of these studies used fictional scenarios and simulated data sets for their demonstrations [P7, P8, P9, P21, P23, P54]. For example, authors in [P23], using a set of test scenarios that simulate five attacks of varying complexity, demonstrate their implementation of smart grid trust visualization. Authors in [P7] show the capabilities of the visualizations using simulated data sets (thus avoiding sensitivity issues). Authors in [P8] explained fictional scenarios that derive and generate data conditions applicable to their visualizations. In other studies [P14, P17, P20, P26, P27, P49], publicly available data sets are used to demonstrate the capabilities of the visualizations.

Eight selected studies (14.8\%) use expert opinions to evaluate their visualizations [P1, P2, P30, P36, P40, P45, P46, P50]. For example, the work reported in [P30] proposed a  set of visual interfaces to help analysts identify and explain off-normal activities. Seven analysts provided feedback on the proposed visualizations by participating in workshops. In [P2], 104 experts, including 12 real users of the system, provided their opinions on the system through a close-ended questionnaire after being exposed to a 3-hour live demonstration of the visualization system.

Three of the selected papers (5.6\%) use academic studies to provide evidence of the proposed CSA visualizations [P3, P39, P47]. For example, in [P3], a two-phase experiment was conducted in a controlled lab environment. The first phase was an observational study to observe how four senior undergraduate students completed a given task in the proposed system based on a given network monitoring data set. The second phase was conducted with seven further participants. Four of them are undergraduate students, two are graduate students, and one is a professional software engineer. In the second phase, participants reviewed the outputs of the first phase and completed a questionnaire about the system.

The maturity of the visualization of six studies (11.1\%) is demonstrated through industrial case studies [P10, P15, P24, P32, P34, P41]. For example, the work reported in  [P32] proposed a platform  for correlating network alerts from disparate logs. Their  prototype was evaluated at the Air Force Research Lab  in New York for one week. It allowed them to  collect perspectives  from analysts and other personnel about the tool's usability and other features that  need to be incorporated into the tool to improve its effectiveness.

Only four studies (7.4\%) [P4, P16, P38, P42] provide evidence of industrial practice for the proposed visualizations. [P4] presents a real-world example of how analysts are using the visualizations at a  large (5000 users)  Security Operations Center (SOC) on a daily basis. In the study, the analysts are defined as experts with experience from 2 to 10 years in network security.  Observations on how analysts use the proposed visualizations were conducted over six months in multiple sessions. They also solicited analyst feedback by email over 12 months. The authors in [P16] explain that the proposed visualization system is used in the real world and have obtained customer feedback. However, the detailed feedback from the customers is not presented in the paper.

\subsection{RQ5: What are the reported challenges in employing visualizations to facilitate cyber security awareness?} \label{sec:challenges}

\begin{table}\caption{Reported challenges} \label{tab:challenges} 
\begin{tabular}{p{0.3\linewidth}p{0.45\linewidth}p{0.1\linewidth}}
\hline
\textbf{Challenge}  & \textbf{Key point} & \textbf{Count}\\ \midrule
Handling large amount of data &   \begin{itemize}
       \item Information cluttering [P2, P3, P4, P6, P7, P12, P13, P15, P16, P18, P21, P25, P28, P30, P31, P44, P50]
      \item Streaming data challenges [P4, P53]
     \end{itemize} &  18\\
Uncertain, missing, or erroneous data & \begin{itemize}
    \item Missing or inadequate information [P2, P10, P49]
    \item Dealing with errors in data [P2, P6, P23, P26]

\end{itemize} & 6 \\
Different data formats and standards & \begin{itemize}
    \item Diverse data sources  [P10, P16, P19, P53]
     \item IP address space [P20, P26]

\end{itemize}  & 6\\
Comprehensibility  of information  &  \begin{itemize}
    \item Visualizations to suit the user and time [P8, P22, P29, P37, P43, P47]
    \item Simple Vs. precise visualizations [P6, P11, P12, P18, P21, P54]
    \item Facilitating the identification of patterns, trends and relationships [P2, P9, P13, P14, P16, P32, P44]
    \item Choosing appropriate aesthetics [P1, P11, P16, P32]
\end{itemize}  & 20\\

Ease of use  &  \begin{itemize}
 \item Easy integration and consistency with existing tools [P11, P25, P28, P53]
 \item Easy to use visualizations  [P14]

\end{itemize} &  5\\
 \bottomrule
\end{tabular}  
\end{table}

\vspace{1em}
\begin{mdframed}
     \begin{itemize}
  \item  We identified several challenges for CSA visualizations reported in the literature.  The most commonly reported challenges are  \textit{handling a large amount of data} and \textit{comprehensibility of information}. 
  
  \item Less commonly reported  challenges are  \textit{uncertain,  missing or   erroneous  data}, \textit{different data formats and standards}, and \textit{ease of use}.
  
\end{itemize}
\end{mdframed}

This section presents the thematic analysis findings for RQ5  and describes various challenges for cyber security visualizations (see Table~\ref{tab:challenges}) that are reported in the selected papers. The data extracted for this section correspond to item D12 in Table 3.

\subsubsection{Handling large amounts of data}
 In the era of big data and the Internet of Things, cyber security data collection volumes are ever-expanding.  As a result, in CSA visualizations, there is a huge degree of complexity involved in storing and viewing a large volume of raw and analyzed data  (33.3\%) [P3, P12, P13, P15, P16, P25].   The streaming nature of data  [P4, P53] can introduce further challenges for analysis due to the continued growth and dynamic nature of the data. Even when information is visualized using several layers, handling large amounts of data is still a huge concern. For example, when historical data is added, the number of layers grows faster, making it difficult to analyze any unfolding trends or patterns. As the density of information increases, users get overloaded with information, and important data could be occluded [P16, P31]. Therefore, it is crucial for CSA visualizations to be flexible and scalable to cater to the immense volumes of data that are generated by modern data sources.

\subsubsection{Uncertain, missing, or erroneous data}

Several studies have discussed challenges with respect to uncertain, missing, or erroneous data for CSA visualizations (11.1\%). Having uncertain, missing, and erroneous data in CSA visualizations means that those visualizations could present misleading information to the user leading to flawed decision making.     
Therefore, CSA visualizations should consider techniques to compensate for data flaws and statistical variability [P26] to deal with false positives [P6] and missing, fragmented or inaccurate data [P2, P49].
       
\subsubsection{Different data formats and standards}   
The number of devices connected and the variety of applications or services employed in current CSA visualizations are very high. It means that creating these visualizations requires a high volume of heterogeneous data formats to be stored and analyzed (11.1\%) [P10, P16, P53]. For example, researchers in [P16] use data from different data sources in their platform for real-time detection and visualization of cyber threats. These data sources are divided into external data and internal data.  External sinkholing, passive DNS, and social media data are external sources examples used in their work. Network flow, logs, and analysis outputs captured inside the network are examples of internal data sources employed in their work.  Having diverse sources and data would require having systems and practices in place to store and analyze apparently uncorrelated data to build effective CSA visualization systems.

\subsubsection{Facilitating comprehension of information}
 Ensuring that users can comprehend and synthesize the provided information is a huge challenge in CSA visualizations. 37.0\%  of primary studies mention this challenge. The CSA visualizations have to be simple enough to enable users to understand the visualization easily and precisely enough to make correct decisions swiftly [P12, P54]. Not all the available information has to be shown to users at once to enable them to make decisions. On the other hand, not providing adequate information could lead to flawed decision making. The information in CSA visualizations should be visualized so that users can quickly and easily identify any patterns, trends, and relationships [P44]. Choosing appropriate aesthetics also plays an important part in facilitating comprehension of the information shown in CSA visualizations [P1]. Another key challenge is providing the correct type of visualization at the right time [P22] and making sure the provided visualizations relate to users' knowledge and experience [P8, P22, P37]. 

\subsubsection{Ease of use}
Several studies (9.3\%) explain that CSA visualizations need to be easy to use in order to be effective and useful.  If visualizations have adequate information for users to make decisions, but the users cannot easily identify or find that information, then those visualizations will not be effective. Since each user will be different, the user requirements have to be considered carefully to understand how to design visualizations that are easy to use. Another common challenge in CSA visualizations is that they are often standalone and do not integrate well with existing tools and data. Users often trust specific tools and data sources they understand and rely heavily upon. So if the CSA visualizations are not consistent with existing tools and systems or do not integrate well, they will be less effective and useful [P53]. When  CSA visualizations do not  integrate well with existing tools  and practices, it limits users' capacity to collaborate, communicate effectively, and share information with others.

\subsection{RQ6: What practices have been reported to implement cyber situational awareness visualizations successfully?} \label{sec:practices}

\vspace{1em}
\begin{mdframed}
     \begin{itemize}
 \item  We identified several practices  to implement the CSA visualizations reported in the literature.  The most commonly reported practices are  \textit{condensed presentations}, \textit{providing context}, and \textit{layouts and aesthetics to reduce visual complexity}. 
  
  \item Less commonly reported  practices are  \textit{flexibility handle differences in data} and \textit{facility to share information}.
\end{itemize}
\end{mdframed}

This section presents the thematic analysis findings for RQ6  and describes key practices (see Table~\ref{tab:practices}) regarding cyber security visualizations reported in the selected papers. The  data extracted for this section correspond to item D13 in Table~\ref{tab:dataextraction}.

\begin{table}[]\caption{\label{tab:practices} Reported practices}
\begin{tabular}{p{0.25\linewidth}p{0.5\linewidth}p{0.1\linewidth}}
\toprule
\textbf{Practice}  & \textbf{Key point} & \textbf{Count}\\ \midrule
Condensed presentation &   \begin{itemize}
    \item Condensed forms of visualizations [P2, P4, P8, P13, P21, P24, P26, P38, P40, P43, P50, P54]
    \item Superimposing different visualizations techniques [P5, P8, P34, P37, P38, P42]
    \item User interactions to support users obtaining information only on demand [P2, P13, P16, P22, P30, P38, P39, P40, P41, P43, P44, P46, P53, P54]

\end{itemize}  & 24
 \\
Providing context  &   \begin{itemize} 
 \item Context-aware adaptive visualizations [P18, P41, P49]
\item Providing trends and patterns [P4, P13, P16, P21, P24, P30, P34, P38, P41, P44, P47, P49, P51, P53]
 \item Details to provide context for threats or risks [P2, P4, P5, P8, P9, P13, P34, P35, P37, P43, P44, P54]
     %  \item Visualizations to provide information on impact of a threat [P2, P8, P9, P13, P35, P43, P54]
     \end{itemize} & 23
     \\
     
 Layouts and aesthetics to reduce visual complexity & \begin{itemize}
    
    \item Reducing complexity of data using layout options [P28, P37, P38, P41, P53].
  \item Providing multiple views for information visualization [P2, P4, P5, P12, P13, P16, P20, P25, P33, P37, P38, P39, P41, P50]
   \item Importance for visual attributes to reduce visual complexity [P1, P2, P3, P5, P8, P11, P13, P16, P22, P37, P43, P46, P50, P53]

\end{itemize} & 23 \\
 
 Facility to  share information  &  
    \begin{itemize}
    \item Providing the ability to share  visualized  information [P3, P4, P10, P12, P16, P37, P38].

\end{itemize} & 7
\\
 Flexibility to handle differences in data &  
    \begin{itemize}
    \item 
    Extra views to handle accurate and missing information [P2, P34, P50].
    \item Flexibility to have different data models [P7].
\end{itemize} & 4\\

 User-driven requirements &  
    \begin{itemize}
    \item Consultation industrial partners or real users or observations to gather requirements [P2, P4, P11, P12, P24, P38, P41, P42, P45].
\end{itemize} & 9\\

 Real world evaluations  &  
     
 \begin{itemize}
 \item Evaluation with real users [P2, P4, P16, P24, P32, P36, P39, P45, P46, P50]
  \item Real-world deployment and short term use [P10, P15, P24, P32, P34, P41]
 \item Real-world deployment and long term use [P4, P16, P38, P42] 
    
\end{itemize}  & 16\\

\bottomrule
%\end{tabular}

\end{tabular}
\end{table}

\subsubsection{Condensed presentation}

CSA information that needs to be visualized is often complex and multi-dimensional. Therefore,  CSA visualization researchers have looked into condensed forms of information representation to provide more information using a single visualization. As detailed in Section~\ref{visualizations}, our primary papers have used various forms of visualization techniques such as \textit{geometrically transformed displays}, \textit{iconic displays}, \textit{dense displays}, \textit{geographic displays}, and \textit{stacked displays}, to present diverse multi-dimensional data in compact ways. Furthermore,   multiple visualization techniques are superimposed to provide additional information to the user in a single visualization. For example, color or shape icons are often used with \textit{geometrically transformed displays}, \textit{geographical displays}, and \textit{stacked displays} to emphasize the status or severity, or impact of particular phenomena (refer to Section~\ref{visualizations}).
Furthermore, user interactions such as \textit{details on demand}, \textit{zooming}, and \textit{filtering}  allow  users to obtain information only on demand, which facilitates showing information in a condensed way.      
        
\subsubsection{Providing context}
As CSA visualizations often deal with a tremendous amount of data,  the user performance in comprehending the provided information and projecting for the future could suffer tremendously without support to reason out the context. In fact, previous research has highlighted that providing context to interpret information is the key to developing CSA [P30]. We observed several ways visualizations available in our primary papers facilitate users to comprehend information by providing context.  For example, researchers in [P4] identify the temporal context of an event as an important design practice for CSA. They used horizon graphs of several flow fields and heatmaps of IP addresses to provide context  to a cyber event.  Furthermore, researchers in [P13] adopt the practice of showing trends and patterns of how the network compromises could affect the organization's performance. Researchers in [P30] attach relevant contextual information  to the charts so that users can easily understand why certain activity changes might be taking place.
On the other hand, limited studies have looked at context-adaptive CSA visualizations.  For example, researchers in [P18]  propose a real-time adaptive system for recommending the appropriate level of detail views tailored for hierarchical network information structures. This system reasons the contextual information associated with the network, user task, and user cognitive load to adapt the network visualization presentation to facilitate context-aware reasoning.

\subsubsection{Layouts and aesthetics to reduce visual complexity}

The visual complexity influences how a user will interact with those visualizations. Several papers have focused on more effective layouts and aesthetics to reduce visual complexity. In terms of having better layouts, the authors of [P28] propose  a top-level layout approach  to perform incremental layout algorithms. This approach allows them to import and display large attack graphs in seconds which previously could take several hours to load.  In [P14], the authors use the client-server layout in Gephi to reduce bipartite graphs' complexities. In [P25], aggregated alert events are presented using multiple coordinated views with timeline, cluster, and swarm model analysis displays. The framework aims to improve situational awareness and to enable an analyst to easily  navigate and analyze large number  of detected events and also be able to combine sophisticated data analysis techniques with interactive visualization for ease of maneuvering through complex information. Researchers in [P4] propose several views to present different types of information. These views include overviews that allow users to scan information within seconds and other views to conduct detailed analysis if needed. Several primary papers discuss the importance of focusing on aesthetics to reduce visual complexity. 
Researchers in [P8] discuss selecting icons/symbols in the visualizations that relate more to the users' day-to-day business. They claim that will allow users to understand and interpret information that is visualized easily. Another paper [P11] discusses using dark background so that users can visualize things unobtrusively in a 3D environment.

\subsubsection{Facility to share information}
Complete CSA is implausible to achieve by considering  interactions  between an individual analyst or decision maker and their technology alone \cite{cooke2013cyber,conti2013towards}.  Achieving complete SA requires diverse stakeholders to collaborate and share information with each other.  Often each stakeholder will have different and sometimes overlapping perspectives  on the situation. Two or more such perspectives will likely need to be combined to obtain complete SA. Unfortunately, there is a lack of technologies conducive to humans collaborating, effectively communicating, and sharing information and knowledge in the context of CSA.  A limited number of our primary papers have reported practices that enable visualization data to be shared with others. For example, researchers in [P4] introduce \emph{watchlists} in their visualizations to manage suspicious IP addresses lists that can be shared with analysts. In [P3], the researchers propose a mind mapping tool that allows analysts to directly
interact with each other and review past analysis, share their findings and divide tasks in a timely manner.

\subsubsection{Flexibility to handle differences and issues in data}
As explained in Section~\ref{sec:challenges}, key challenges for CSA visualizations include handling differences in data formats and standards, and dealing with uncertain and erroneous data.   A limited number of primary papers in this SLR report practice handling these differences and data issues. For example, researchers in [P7] explain previous graph-based tools that focus on specific analytic use cases against fixed data models and propose a schema-free data model to decouple from the storage implementation. The proposed approach applies data transformations that map source data elements to nodes, edges, and their properties rather than relying on a fixed schema for the data sources. Researchers in [P2] propose a method to deal with possible missing or inaccurate information in alert messages. Their algorithms consider two different matches: i)~approximate matches and ii)~exact matches.  The exact match allows taking into account possible inaccurate or wrong information, which includes but is not limited to a missing source IP address in the alert and a mismatch in the CVE due to different classifications used by the underline IDS.

\subsubsection{User driven requirements}
A clear understanding of user needs  is an essential part of software design and could be considered  one of the deciding factors of the success of systems \cite{maguire2002user}.  However, we only observed that 16.7\% of the primary papers had consulted industrial partners or real users when designing CSA visualizations. For example, researchers in [P24] conduct a series of brainstorming and interviews with analysts, network managers, security researchers, and visualization researchers before coming up with visualization mock-ups to facilitate immediate high-level SA. Furthermore, researchers in [P12] visit and observe four Security Operations Centers (SOC) of their industrial partners to understand cyber security collaborative practices before designing a collaborative  3D Cyber Common Operating Picture Platform.

\subsubsection{Real world evaluations}
 The software engineering research community had emphasized the criticality of rigorous evaluation to assess the appropriateness of the proposed solutions \cite{zannier2006success}. However, as detailed in Section~\ref{maturity}, among the selected studies,  there is a lack of rigorous evaluation that utilizes more mature methods such as real-world deployments and case studies with real users. Our findings clearly demonstrate that most primary papers do not involve real users in their evaluations. Only a few papers looked into conducting case studies or deploying the proposed visualization systems in the real world to understand how the users perceive those systems in practice.

\section{Discussion} \label{sec:discussion} 

There is an increasing realization that cyber security visualizations can enable significant progress towards achieving the goal of CSA. Throughout this review, we have identified, categorized, and discussed the knowledge related to  CSA visualizations in various dimensions.  
This section will summarize the key findings from this SLR (see Section \ref{key_insights}) and discuss the potential future research and development opportunities in the CSA visualization domain based on the identified key limitations and gaps (see Section \ref{future}). 

\subsection{Key insights} \label{key_insights}

\textbf{Focus on operational-level staff:} 
We identified several stakeholders who use and benefit from CSA visualizations (see Section~\ref{sec:stakeholders}).  Our results clearly show that most papers (64.8\%) provide visualizations for operational-level staff such as network analysts, risk analysts, and security analysts. 
From an organizational perspective, there is an evident lack of scientific research that presents information for \textit{managers and higher-level decision makers}.  Usually, \textit{managers and higher-level decision makers} are tasked with overseeing the operations and activities of an organization and making strategic decisions that can influence the future of the organization.  Often they lack cyber expertise \cite{james2018making}; hence in the absence of CSA visualizations, they may have to rely on domain experts to interpret the cyber security status of the organizations, causing delays in the decision making process.  
Outside the organizational context, we found only two studies that provide CSA visualizations targeted at \textit{non-expert users}.  With the ever-increasing security threats online and lack of cyber security awareness of non-expert users who act in cyberspace,  it is alarming to have such a  limited number of studies with CSA visualizations targeted at \textit{non-expert users}.

\textbf{Limited attention to external data sources:}
Our analyses provide important insights into the types of data sources used by CSA visualizations. 
We found that several studies reported difficulties dealing with diverse data sources, and in particular, our results indicate that \textit{external data sources} are the least common source of CSA visualizations.
This is concerning since being limited to internal cyber security data and knowledge could limit situational understanding of cyber security threats and risks, hindering the effectiveness of cyber security decision making \cite{Petrenko2020, Czejdo2014}.

 \textbf{Limited attention to specific information types:}
 Our results reveal diverse types of information visualized through the CSA visualizations. The most common type of information visualized is the \textit{threat information} (55.6\%). However, we observed a lack of attention to visualizing \textit{impact information} and \textit{response plans}.   Understanding the business impact of a cyber incident and response plans allows  effective management of cyber risks and more targeted responses to cyber incidents \cite{musman2011computing}. Furthermore, \textit{shared information} is the least common type of information visualized in CSA visualizations. Given that  previous research has highlighted that team-level SA is of utmost importance for complete CSA and communication and information coordination is at the heart of team-level situation awareness \cite{cooke2013cyber}, lack of  visualizations to  support  communication  and collaboration among different team members is  concerning.

\textbf{Lack of attention to several  CSA visualization techniques and  user interactions:}  We categorized the visualizations of the selected papers under nine visualization techniques. From the results presented in Section~\ref{ss:visualization techniques}, it is clear that  \textit{iconic displays} and \textit{geometrically transformed displays} are the most popular visualizations techniques used in the studies. \textit{Iconic display} is an interesting way to encode information, while increasing the hedonic quality of the visualizations. \textit{Geometrically transformed displays}   allow users to understand complex, multi-dimensional cyber data through interesting transformations. Other visualization techniques (e.g., \textit{immersive environments}) are less employed in the selected set of papers. It was also clear that many visualizations combine multiple visualization techniques, often by superimposing them, to provide more information in a condensed manner. However, more user evaluations are needed to comment on their effectiveness.
The power of visualization can be enhanced through user interactions. However, we noticed that a significant number of papers (16 papers, 29.6\%) did not discuss user interactions. As explained in Section~\ref{ss:visualization techniques}, it is uncertain whether the authors do not emphasize these features or visualizations do not have user interactions. Furthermore, we noticed that while user interactions like \textit{zooming}, \textit{filtering}, and \textit{details on demand} have gained much attention, other user interactions have gained less attention. For example,  \textit{extract/share} and \textit{move/rotate} were only found in four papers respectively, and \textit{linking/brushing} was only found in one study. 
	
\textbf{Facilitating  only lower-level of CSA:}
Understanding what is happening in the cyber environment is only the first level of CSA (i.e., \textit{perception} level as described in  Section~\ref{background:CSA}). The ability to comprehend and interpret the current cyber situation is  crucial to move beyond \textit{perception} level and reach \textit{comprehension} level.  Several studies in the SLR report challenges comprehension of the information visualized (see Section~\ref{sec:challenges}). Our results  show that most studies   (92.6\%) facilitate  the \textit{perception} level, compared with only 53.7\% of the studies that facilitate up to the \textit{comprehension} level.  Unfortunately, not having the ability to understand the data and its relationships could lead to poor interpretation of the displayed information and hence could reduce the power of visualizations. 
As discussed in Section~\ref{background:CSA} and Section~\ref{CSAsupport}, to move beyond the \textit{comprehension} level to \textit{projection} level, users should also be able to identify the future state of threats and possible future actions. Unfortunately, our results provide evidence that the \textit{projection} level is the least supported through the CSA visualizations in the selected papers.

\textbf{Lack of rigorous evaluations:}
As explained in Section~\ref{maturity}, most studies either do not provide evidence or only provide demonstrations/toy examples as evidence of the proposed visualizations. Lack of rigorous evaluation could be the main reason for the limited number of studies (7.4\%) that provide evidence for industrial practice.

\textbf{Mapping between challenges and practices:} 
Figure~\ref{fig:mapping} presents a mapping of the identified challenges in Section~\ref{sec:challenges} onto the practices reported in Section~\ref{sec:practices}. 
First, this mapping provides readers (both researchers and practitioners) with a quick way to identify the relationships between challenges (i.e., \textit{exacerbation}). For example, having a large amount of data could result in challenges for the comprehensibility of the visualized information and could hinder the ease of use of the CSA visualizations.
Second, this mapping provides readers with a way to identify how practices could help overcome the challenges in CSA visualizations (i.e., \textit{support}). For example, driving CSA visualization designs based on user needs and preferences, focusing on better layouts and aesthetics to reduce visual complexity, and providing the ability for users to share visualized information easily could allow CSA visualization to be easy to use. Furthermore, conducting real-world evaluations will ultimately provide evidence of whether the designed and developed CSA visualizations are easy to use in practice. In summary, the mapping in Figure~\ref{fig:mapping} provides anyone interested in CSA visualizations with the ability to understand the challenge space better and in more detail, and how changed practices could alleviate these challenges.

\Figure(topskip=0pt, botskip=0pt, midskip=0pt) [width=0.999\linewidth] {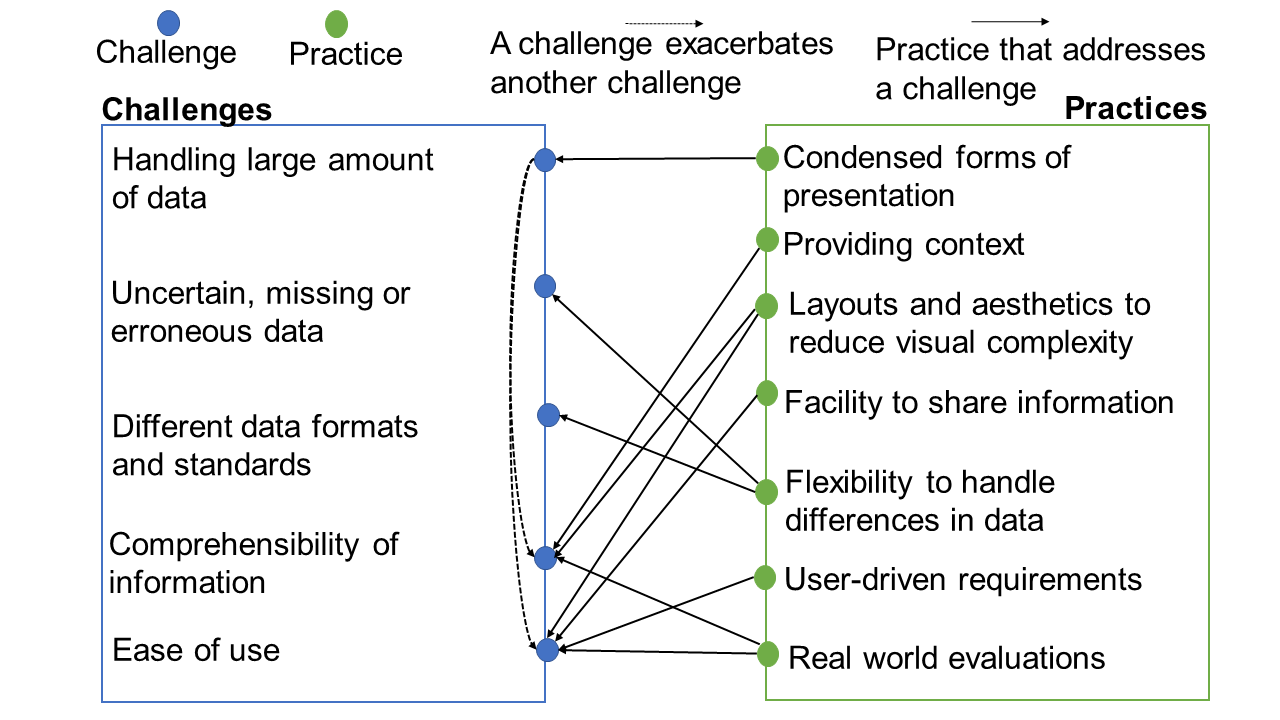}
{Mapping of challenges to practices.\label{fig:mapping}}

\subsection{Recommendations for future research} \label{future}

\textbf{CSA visualizations for higher-level decision makers and non-expert users:}  To be best placed to make cyber security decisions effectively and efficiently higher-level decision makers, including executives, should have access to information on the entire organization’s potential cyber security risks, opportunities, and challenges in a format that is easy to digest and translate to the business dimensions. Hence future research could be conducted to specifically target the design of CSA visualizations for managers and higher-level decision makers. A better understanding of their information needs and visualization preferences will facilitate the development of effective visualizations for this cohort. Outside the organizational context, there are opportunities to design visualizations for CSA focusing on non-expert users. It can be expected that such visualization support will help non-expert users be proactive about their online safety. 

\textbf{Cyber Common Operating picture focusing on all levels of staff:} Future studies can invest effort in developing fully customizable Common Operating Pictures to facilitate cyber security decision making \cite{conti2013towards}. We anticipate such platforms will combine data from various data sources such as  Security Information and Event Management (SIEM) systems, Intrusion Detection Systems (IDS), logs, data from security training and awareness programs, patching coverage, new critical vulnerabilities, and external sources of threat intelligence,  to provide all levels of staff a common view of cyberspace to facilitate collective decision making. Among our selected primary papers, we found only a few papers [P5, P8, P12, P20] that explicitly claim to focus on Common Operating Pictures. These studies are still in their infancy and cannot provide the true power of Cyber Common Operating Pictures. These systems further lack fully customizable dashboards that allow organizations to adapt the information they want to visualize and tailor visualizations to a particular audience.

\textbf{Context-aware adaptive visualizations:} The ultimate goal of CSA  visualizations should be to get the correct information to the right person, at the right time, in the right way to facilitate swift decision making. Unfortunately, the sheer volume of cyber security data could lead to over-crowding of displays, decreasing the power of visualizations and decreasing the capacity for a human to identify key information, trends, and data patterns. As explained in Section~\ref{sec:practices}, condensed or summarised forms of information visualizations and powerful user interactions could allow users to find and navigate to the appropriate level of detail. However, this approach places control on the user to identify and navigate where they need to focus on making decisions. Manual navigation of the required information could be seen as a laborious, error-prone process that could create a cognitive overload for users. Therefore, we argue that future research should focus more on visualizations capable of automatically adapting the information and visualization techniques used based on the context, user needs, and task at hand. Only a few papers [P18, P41, P49] in this SLR discussed this concept; hence we believe there is a clear gap in this area and argue that this research area (i.e., adaptive and context-aware visualization) should be investigated further in the future.

\textbf{Novel data sources and data source agnostic visualizations:}
Future researchers can explore further into how external data sources such as Open-Source Intelligence (OSINT) can be integrated into CSA visualizations.   OSINT can be considered an early warning source for cyber security events such as vulnerability exploits \cite{Le2019}. For example, publicly available data sources such as Twitter could be used to identify emerging threats and cyber attacks. We believe that combining internal and external data sources could result in a better CSA for the organizations.

As identified by several papers in Section~\ref{sec:challenges}, introducing new data sources and fusing diverse data sources can create additional challenges for CSA visualizations.  These findings suggest that future CSA visualizations should give careful consideration to heterogeneous data types which need to be conveniently stored and prepared for analysis \cite{10.1145/3230833.3232798}.  In this context, we emphasize that it is important to follow the principles and best practices of scalable big-data systems to store and analyze such heterogeneous data for building effective CSA visualizations.
Organizations have different environments and data sources that contribute to their own CSA, so there are no one-size-fits-all visualization solutions. Therefore, researchers could invest their efforts in automated CSA visualization generation from machine-readable data models \cite{Komarkova2018}. 
Furthermore, AI-based techniques can then be applied to recommend visualization models based on data sources provided by an organization \cite{Coulter2020}. The automated CSA dashboard generation will be beneficial for organizations to monitor key information from the data sources dynamically, fast prototype their organizational CSA knowledge, and validate their CSA design.

 \textbf{Facilitating collaboration and information sharing:}
Achieving complete SA requires members of different teams and different organizational positions, working across different work shifts to collaborate and share information with each other ~\cite{cooke2013cyber,6725797, rajivan2017impact}. Lack of the ability to collaborate and share information within the organization could limit the ability of organizations to take full advantage of their staff's expertise and relationships for the management of vulnerabilities, threats, and incidents, as well as other cyber security activities. 
We only found a limited number of visualizations in this SLR that provided some form of support for collaboration and information sharing. Therefore, we assert that collaboration and information sharing should be considered an integral part of CSA visualizations in the future. Collaboration and information sharing should be considered within the organization and across organizations. More prominence can be given to visualizations that facilitate collective decision making, sharing of information within different applications of the same organization, and sharing of information within organizations. 

\textbf{User-centered design approaches:} We assert that user-centered design should be an intrinsic part of the design philosophy of CSA visualizations. Traditional practices of user-centered design incorporate a clear understanding of users' needs, wants, and limitations throughout the design process, which help evaluate the effectiveness of the proposed systems or tools \cite{noyes1999user}. Therefore, we emphasize that first understanding users' CSA needs and then iteratively improving the visualizations based on their feedback is crucial to implementing usable and effective visualization. However, only nine studies (16.7\%) in this SLR have attempted to understand the requirements from users for CSA visualizations. Furthermore, only one study [P30] discusses iterative user involvement throughout the process, including brainstorming, design, and evaluation. We believe the availability and cost of experts could also create challenges for user-centered design approaches in the CSA visualization domain. Therefore, we assert that adopting user-centered design approaches within the cyber security visualization domain requires effective and efficient research methods that facilitate user involvement.
 
\textbf{Visualization support to project future events, consequences, and possible actions:} We emphasize the need for a gradual but inevitable transition of future visualization approaches towards facilitating the \textit{projection} level to achieve comprehensive CSA.   We anticipate that complex data analysis approaches, which may stem from AI and ML techniques, may be used in this regard \cite{husak2018survey}.

\textbf{More rigorous real-world evaluations:}
Rigorous real-world evaluations will improve the applicability and quality of the research outcomes \cite{zannier2006success}. However, this SLR observed that only a small percentage of the studies had been evaluated in an industrial setting, which may be due to fewer industry collaborations. Therefore, future researchers should pay more attention to rigorously evaluating the CSA visualizations using approaches with industrial relevance. It will lead to more practical and usable research outcomes.

\section{Threats to validity} \label{threats}
We followed the guidelines provided by \cite{kitchenham2007guidelines} strictly; however, we had similarities to other SLRs regarding validity threats, which we will discuss below.

\textbf{Missing primary studies:} Most of the SLRs face the limitation of missing primary
studies. It is mainly due to limitations in the search method and non-comprehensive venues. To minimize the effects of this issue, we used several strategies. We used Scopus as our search engine. Scopus is the largest indexing system leading to the most comprehensive search results among other digital libraries~\cite{zahedi2016systematic}, allowing us to expand the coverage of relevant studies.   Furthermore, our search string was carefully identified. When constructing the search terms, we consulted the search strings used in the existing SLRs~\cite{Franke2014}.   We iteratively improved the search string based on the pilot searches by ensuring all known papers could be captured through the search string. All authors carefully checked the search string before executing the search.   
Furthermore, although we did not impose any restrictions on the publication date of the papers, we acknowledge that the studies added to the database after the search date (\textit{i.e.,} February 2021) are not considered in the review, which is an inevitable limitation in SLR studies~\cite{tran2021integrating}.  

\textbf{Bias in study selection:} 
Studies can be selected based on the subjective judgments of researchers regarding whether or not they meet the selection criteria.
We strictly followed the predefined review protocol to address this issue, recording the exclusion reasons for all excluded papers. In addition, a pilot set of selected studies was shared with all the authors to make sure all authors agreed with the inclusion and exclusion criteria. The first three authors largely conducted the study selection and had ongoing internal discussions about any papers that raised doubts about their inclusion or exclusion decisions; the remaining authors were consulted whenever a decision could not be made.

\textbf{Bias in data extraction and analysis:} 
To reduce the bias in the data extraction,  we first created a data extraction form (see Table~\ref{tab:dataextraction}) to extract and analyze the data consistently in order to answer the RQs of this SLR.
Then, the first three authors conducted a data extraction pilot with a subset of papers. Any differences in the data extraction were discussed and resolved; where necessary, the remaining authors were consulted. After that, the papers were divided, and the first three authors extracted data separately. To analyze the extracted data, both quantitative and qualitative methods were applied. It should be noted that we did not have any interpretation unless the data items were explicitly provided by the study. It should be noted that occasionally it was difficult to interpret the extracted data because of a lack of sufficient information about the data items. Consequently, in some instances, interpretation and analysis of the data were subjective, which might have influenced the data extraction results.

\section{Conclusion} \label{sec:conclusion}

With cyber attacks becoming ever more sophisticated and creating potentially disruptive impacts, underadjusting the cyber security landscape is more necessary than ever. A picture is worth a thousand words; hence cyber security visualizations play a pivotal role in conveying complex cyber security information efficiently and effectively. 
This paper reports our research efforts to systematically review the literature on the CSA visualizations and shed light on important aspects of this emerging research field. 

We have conducted rigorous analysis and systematic synthesis of 54 papers reporting research on CSA visualizations. Our research questions systematize and learn different stakeholders of CSA visualizations, different information types visualized, data sources employed, visualization techniques used, CSA levels that can be achieved through the proposed visualizations, the maturity of the proposed visualizations, challenges identified in designing and developing CSA visualizations,  and practices that have been reported to implement CSA visualization successfully.

The findings of this SLR will help to inform researchers and practitioners of the main limitations and barriers to the design, development, and adoption of CSA visualizations and help direct future research in this area. For example, we found a lack of research focused on higher-level decision makers, and non-expert users. Our results also reveal that most visualizations do not reach the required level of maturity. Finally, we also provided a  mapping between challenges and practices, and we anticipate this will be beneficial for researchers and CSA designers so they can more easily understand what practices exist for facilitating each challenge reported for CSA visualizations.

We have provided guidance for the areas of future research through 8 recommendations. Furthermore, while we acknowledge that our study does not provide a complete view of the CSA, the visualizations are one of the most important components of any technological system designed for CSA. Hence, we take the first step to highlight this area of research and lay the foundation for developing effective CSA technologies and systems in the future.

\section*{Acknowledgment}
We thank Dr. Praveen Gauravaram from Tata Consultancy Services Limited (TCS) for providing detailed feedback on earlier versions of this SLR.

\typeout{}

\bibliographystyle{IEEEtran}
\bibliography{access}

\begin{table*}[b!]
\begin{appendices}
\section{Selected primary studies}

\end{appendices}
\begin{tabular}{p{0.03\textwidth}p{0.4\textwidth}p{0.2\textwidth}p{0.2\textwidth}p{0.04\textwidth}}
\addlinespace
\midrule
ID & Title                                                                                       & Author(s)                                               & Venue                                                   & Year \\ \midrule
P1  & What makes for effective visualization in cyber situational   awareness for non-expert users?                                       & Carroll F., Chakof A., Legg P.                                                                              &  International Conference on Cyber Situational Awareness                                                                                                                                                                                & 2019 \\\addlinespace
P2  & MAD: A visual analytics solution for Multi-step cyber Attacks   Detection                                                           & Angelini M., Bonomi S., Lenti S., Santucci G., Taggi S.                                                     & Journal of Computer Languages                                                                                                                                                                                                                                                               & 2019 \\\addlinespace
P3  & AOH-Map: A Mind Mapping System for Supporting Collaborative   Cyber Security Analysis                                               & Zhong C., Alnusair A., Sayger B., Troxell A., Yao J.                                                        & IEEE Conference on Cognitive and Computational Aspects of Situation Management                                                                                                                                                                          & 2019 \\\addlinespace
P4  & Situ: Identifying and explaining suspicious behavior in   networks                                                                  & Goodall J.R., Ragan E.D., Steed C.A., Reed J.W., Richardson   G.D., Huffer K.M.T., Bridges R.A., Laska J.A. & IEEE Transactions on Visualization and Computer Graphics                                                                                                                                                                                                                                    & 2019 \\\addlinespace
P5  & Cyber kill chain based threat taxonomy and its application on   cyber common operational picture                                    & Clio S., Han I., Jeong H., Kim J., Koo S., Oh H., Park M.                                                   &  International Conference on Cyber Situational Awareness                                                                                                                                                                & 2018 \\\addlinespace
P6  & Combining real-time risk visualization and anomaly detection                                                                        & Väisänen T., Noponen S., Latvala O.-M., Kuusijärvi J.                                                       & ACM International Conference Proceeding Series                                                                                                                                                                                                                                              & 2018 \\\addlinespace
P7  & Mission-focused cyber situational understanding via graph   analytics                                                               & Noel S., Rowe P.D., Purdy S., Limiero M., Lu T., Mathews W.                                                 & International Conference on Cyber Conflict                                                                                                                                                                                                                                        & 2018 \\\addlinespace
P8  & A comparative analysis of visualization techniques to achieve   cyber situational awareness in the military                         & Llopis S., Hingant J., Perez I., Esteve M., Carvajal F., Mees   W., Debatty T.                              & International Conference on Military Communications and Information Systems                                                                                                                                                                                           & 2018 \\\addlinespace
P9  & Comparative analysis and patch optimization using the cyber   security analytics framework                                          & Abraham S., Nair S.                                                                                         & Journal of Defense Modeling and Simulation                                                                                                                                                                                                                                                  & 2018 \\\addlinespace
P10 & Blue Team Communication and Reporting for Enhancing   Situational Awareness from White Team Perspective in Cyber Security Exercises & Kokkonen T., Puuska S.                                                                                      & Conference on Internet of Things and Smart Spaces                                                                                                                                                  & 2018 \\\addlinespace
P11 & Enhancing cyber defense situational awareness using 3D   visualizations                                                             & Kullman K., Cowley J., Ben-Asher N.                                                                         & International Conference on Cyber   Warfare and Security                                                                                                                                                                                          & 2018 \\\addlinespace
P12 & From Cyber Security Activities to Collaborative Virtual   Environments Practices Through the 3D CyberCOP Platform                   & Kabil A., Duval T., Cuppens N., Le Comte G., Halgand Y.,   Ponchel C.                                       & International Conference on Information Systems Security                                                                                                                                                      & 2018 \\\addlinespace
P13 & Cyber situational awareness: from geographical alerts to   high-level management                                                    & Angelini M., Santucci G.                                                                                    & Journal of Visualization                                                                                                                                                                                                                                                                    & 2017 \\\addlinespace
P14 & Deriving cyber use cases from graph projections of cyber data   represented as bipartite graphs                                     & Eslami M., Zheng G., Eramian H., Levchuk G.                                                                 & IEEE International Conference on Big Data                                                                                                                                                                                                        & 2017 \\\addlinespace
P15 & A Study into Detecting Anomalous Behaviours within HealthCare   Infrastructures                                                     & Boddy A., Hurst W., Mackay M., El Rhalibi A.                                                                & International Conference on   Developments in eSystems Engineering                                                                                                                                                                                   & 2017 \\\addlinespace
P16 & OwlSight: Platform for real-time detection and visualization   of cyber threats                                                     & Carvalho V.S., Polidoro M.J., Magalhaes J.P.                                                                & IEEE International Conference on Big Data   Security on Cloud & 2016 \\\addlinespace
P17 & An expert system for facilitating an institutional risk   profile definition for cyber situational awareness                        & Graf R., Gordea S., Ryan H.M., Houzanme T.                                                                  & ICISSP 2016 - International Conference  on Information Systems Security and Privacy                                                                                                                                                                                 & 2016 \\\addlinespace
P18 & Adaptive visualization of complex networks with focalpoint: A   context aware level of details recommender system                   & Inibhunu C., Langevin S.                                                                                    & Proceedings of the Human Factors and Ergonomics Society                                                                                                                                                                                                                                     & 2016 \\\addlinespace
P19 & Web-Based Smart Grid Network Analytics Framework                                                                                    & Pietrowicz S., Falchuk B., Kolarov A., Naidu A.                                                             & IEEEInternational Conference on   Information Reuse and Integration                                                                                                                                                                                      & 2015 \\\addlinespace
P20 & Configurable IP-space maps for large-scale, multi-source   network data visual analysis and correlation                             & Miserendino S., Maynard C., Freeman W.                                                                      & Proceedings of SPIE - The International Society for Optical   Engineering                                                                                                                                                                                                                   & 2014 \\\addlinespace
P21 & Visualization of security metrics for cyber situation   awareness                                                                   & Kotenko I., Novikova E.                                                                                     & International Conference on Availability,   Reliability and Security                                                                                                                                                                                    & 2014 \\\addlinespace
P22 & CyberVis: Visualizing the potential impact of cyber attacks on   the wider enterprise                                               & Creese S., Goldsmith M., Moffat N., Happa J., Agrafiotis I.                                                 & IEEE International Conference on Technologies for   Homeland Security                                                                                                                                                                                                     & 2013 \\\addlinespace
\bottomrule
\end{tabular}
\end{table*}

\clearpage

\begin{table*}[h!]
\begin{tabular}{p{0.03\textwidth}p{0.4\textwidth}p{0.2\textwidth}p{0.2\textwidth}p{0.04\textwidth}}
\midrule
ID & Title                                                                                       & Author(s)                                               & Venue                                                   & Year \\ \midrule

P23 & CyberSAVe - Situational awareness visualization for cyber   security of smart grid systems                                          & Matuszak W.J., DiPippo L., Sun Y.L.                                                                         & ACM International Conference Proceeding Series                                                                                                                                                                                                                                              & 2013 \\\addlinespace
P24 & Visualization design for immediate high-level situational   assessment                                                              & Erbacher R.F.                                                                                               & ACM International Conference Proceeding Series                                                                                                                                                                                                                                              & 2012 \\\addlinespace
P25 & Visualization techniques for computer network defense                                                                               & Beaver J.M., Steed C.A., Patton R.M., Cui X., Schultz M.                                                    & Proceedings of SPIE - The International Society for Optical   Engineering                                                                                                                                                                                                                   & 2011 \\\addlinespace
P26 & EMBER: A global perspective on extreme malicious behavior                                                                           & Yu T., Lippmann R., Riordan J., Boyer S.                                                                    & ACM International Conference Proceeding Series                                                                                                                                                                                                                                              & 2010 \\\addlinespace
P27 & Intrusion monitoring in process control systems                                                                                     & Valdes A., Cheung S.                                                                                        & Annual Hawaii International Conference   on System Sciences                                                                                                                                                                                                & 2009 \\\addlinespace
P28 & A graph-theoretic visualization approach to network risk   analysis                                                                 & O'Hare S., Noel S., Prole K.                                                                                & International Workshop on Visualization for Computer Security                                                                                                                                                      & 2008 \\\addlinespace
P29 & Cyberspace situation representation based on Niche Theory                                                                           & Zhuo Y., Zhang Q., Gong Z.                                                                                  & IEEE International Conference on   Information and Automation                                                                                                                                                                                        & 2008 \\\addlinespace
P30 & Putting security in context: Visual correlation of network   activity with real-world information                                   & Pike W.A., Scherrer C., Zabriskie S.                                                                        &  Proceedings of the Workshop on Visualization for   Computer Security                                                                                                                                                                                                          & 2008 \\\addlinespace
P31 & Visualizing cascading failures in critical cyber   infrastructures                                                                  & Kopylec J., D'Amico A., Goodall J.                                                                          & IFIP Advances in Information and Communication Technology                                                                                                                                                                                                                                   & 2008 \\\addlinespace
P32 & A visualization paradigm for network intrusion detection                                                                            & Livnat Y., Agutter J., Moon S., Erbacher R.F., Foresti S.                                                   & Annual IEEE System, Man and   Cybernetics Information Assurance Workshop                                                                                                                                                                       & 2005 \\\addlinespace
P33 & Visualization as an aid for assessing the mission impact of   information security breaches                                         & D'Amico A., Salas S.                                                                                        & Information Survivability Conference and   Exposition                                                                                                                                                                                                  & 2003 \\\addlinespace
P34 & ML-based data anomaly mitigation and cyber-power transmission   resiliency analysis                                                 & Anshuman Z.N., Sajan K.S., Srivastava A.K.                                                                  & IEEE International Conference on Communications, Control,   and Computing Technologies for Smart Grids                                                                                                                                                          & 2020 \\\addlinespace
P35 & A novel architecture for attack-resilient wide-area protection   and control system in smart grid                                   & Singh V.K., Govindarasu M.                                                                                  &  Resilience Week                                                                                                                                                                                       & 2020 \\\addlinespace
P36 & Understanding and Enabling Tactical Situational Awareness in a   Security Operations Center                                         & Mullins R., Nargi B., Fouse A.                                                                              & Advances in Intelligent Systems and Computing                                                                                                                                                                                                                                               & 2020 \\\addlinespace
P37 & A Dynamic Visualization Platform for Operational Maritime   Cybersecurity                                                           & Zhao H., Silverajan B.                                                                                      & International Conference on Cooperative Design, Visualization and Engineering                                                                                                                                             & 2020 \\\addlinespace
P38 & Insight2: A modular visual analysis platform for network   situational awareness in large-scale networks                            & Kodituwakku H.A.D.E., Keller A., Gregor J.                                                                  & Electronics (Switzerland)                                                                                                                                                                                                                                                                   & 2020 \\\addlinespace
P39 & Alert characterization by non-expert users in a cybersecurity   virtual environment: A usability study                              & Kabil A., Duval T., Cuppens N.                                                                              & International Conference on Augmented Reality, Virtual Reality and Computer Graphics                                                                                                                                       & 2020 \\\addlinespace
P40 & Operator impressions of 3d visualizations for cybersecurity   analysts                                                              & Kullman K., Asher N.B., Sample C.                                                                           & European Conference on Information Warfare and Security                                                                                                                                                                                                                           & 2019 \\\addlinespace
P41 & A Tri-Modular Human-on-the-Loop Framework for Intelligent   Smart Grid Cyber-Attack Visualization                                   & Sundararajan A., Khan T., Aburub H., Sarwat A.I., Rahman S.                                                 & Conference Proceedings - IEEE SOUTHEASTCON                                                                                                                                                                                                                                                  & 2018 \\\addlinespace
P42 & DiPot: A distributed industrial honeypot system                                                                                     & Cao J., Li W., Li J., Li B.                                                                                 & International Conference on Smart Computing and Communication                                                                                                                                                    & 2018 \\\addlinespace
P43 & Dagger: Modeling and visualization for mission impact   situation awareness                                                         & Peterson E.                                                                                                 & IEEE Military Communications Conference                                                                                                                                                                                                                                & 2016 \\\addlinespace
P44 & Enhancing cyber situation awareness for Non-Expert Users using   visual analytics                                                   & Legg P.A.                                                                                                   & International Conference on Cyber Situational Awareness,   Data Analytics and Assessment                                                                                                & 2016 \\\addlinespace
P45 & Results and lessons learned from a user study of display   effectiveness with experienced cyber security network analysts           & Garneau C.J., Erbacher R.F., Etoty R.E., Hutchinson S.E.                                                    & Learning from Authoritative Security   Experiment Results                                                                                                                                                                                                             & 2016 \\\addlinespace

\bottomrule

\end{tabular}
\end{table*}

\begin{table*}[th!]
\begin{tabular}{p{0.03\textwidth}p{0.4\textwidth}p{0.2\textwidth}p{0.2\textwidth}p{0.04\textwidth}}
\midrule
ID & Title                                                                                       & Author(s)                                               & Venue                                                   & Year \\ \midrule
P46 & Visual analytics for cyber red teaming                                                                                              & Yuen J., Turnbull B., Hernandez J.                                                                          & IEEE Symposium on Visualization for Cyber Security                                                                                                                                                                                                                & 2015 \\\addlinespace

P47 & Ensemble visualization for cyber situation awareness of   network security data                                                     & Hao L., Healey C.G., Hutchinson S.E.                                                                        & IEEE Symposium on Visualization for Cyber Security                                                                                                                                                                                                                  & 2015 \\\addlinespace
P48 & Towards an integrated defense system for cyber security   situation awareness experiment                                            & Zhang H., Wei S., Ge L., Shen D., Yu W., Blasch E.P., Pham   K.D., Chen G.                                  & The International Society for Optical   Engineering                                                                                                                                                                                                                   & 2015 \\\addlinespace
P49 & Visual structures for seeing cyber policy strategies                                                                                & Stoll J., Bengez R.Z.                                                                                       & International Conference on Cyber Conflict                                                                                                                                                                                                                                       & 2015 \\\addlinespace
P50 & VAFLE: Visual analytics of firewall log events                                                                                      & Ghoniem M., Shurkhovetskyy G., Bahey A., Otjacques B.                                                       & The International Society for Optical   Engineering                                                                                                                                                                                                                   & 2014 \\\addlinespace
P51 & Capturing human cognition in cyber-security simulations with   NETS                                                                 & Giacobe N.A., McNeese M.D., Mancuso V.F., Minotra D.                                                        & IEEE International Conference on   Intelligence and Security Informatics: Big Data, Emergent Threats, and   Decision-Making in Security Informatics                                                                                                                    & 2013 \\\addlinespace
P52 & On detection and visualization techniques for cyber security   situation awareness                                                  & Yu W., Wei S., Shen D., Blowers M., Blasch E.P., Pham K.D.,   Chen G., Zhang H., Lu C.                      & The International Society for Optical   Engineering                                                                                                                                                                                                                   & 2013 \\\addlinespace
P53 & Visualization for cyber security command and control                                                                                & Langton J.T., Newey B., Havig P.R.                                                                          & The International Society for Optical   Engineering                                                                                                                                                                                                                   & 2010 \\\addlinespace
P54 & ViSAw: Visualizing threat and impact assessment for enhanced   situation awareness                                                  & Nusinov M., Yang S.J., Holsopple J.                                                                         &IEEE Military Communications Conference                                                                                                                                                                                                                                 & 2009
\\\bottomrule

\end{tabular}
\end{table*}

\clearpage

\begin{IEEEbiography}[{\includegraphics[width=1in,height=1.25in,clip,keepaspectratio]{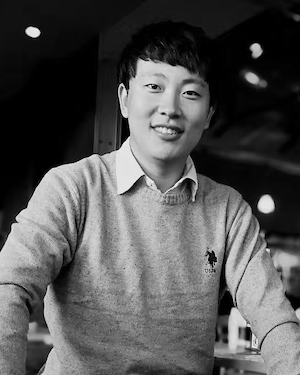}}]{LIUYUE JIANG} received the M.S. degree in computer science from School of Computer Science, The University of Adelaide, Australia, in 2019, where he is currently pursuing the Ph.D. degree with Centre for Research on Engineering Software Technologies (CREST) at the University of Adelaide, and he is funded by the Cyber Security Cooperative Research Centre (CSCRC).
\end{IEEEbiography}

\begin{IEEEbiography}[{\includegraphics[width=1in,height=1.25in,clip,keepaspectratio]{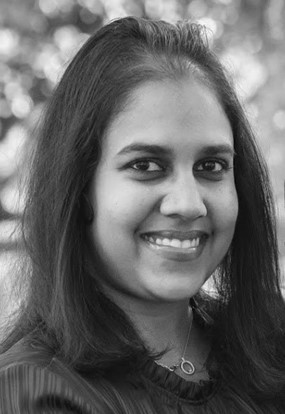}}] 
{ASANGI JAYATILAKA} is a post-doctoral researcher at the Centre for Research on Software Technologies (CREST) at the University of Adelaide (UoA). Prior to joining CREST, she worked as a lecturer in Software Engineering and Computer Science at the UoA, School of Computer Science. She received her PhD from the School of Computer Science at UoA. She is passionate about research on human aspects in computing. This includes studying the effects of different human aspects on technology development, whether and/or to what extent these are accounted for and how we can best use these to build better tools and technologies that are both usable and effective. She has extensive experience in both qualitative and quantitative research methods. Her work has led to designing, implementing, and evaluating technologies and tools in various domains, including cyber security, digital health, and pervasive computing.
\end{IEEEbiography}

\begin{IEEEbiography}[{\includegraphics[width=1in,height=1.25in,clip,keepaspectratio]{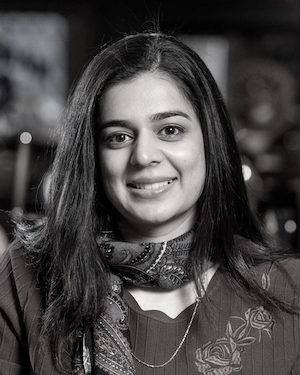}}]{MEHWISH NASIM} is a lecturer in Computing and Mathematical Sciences at the College of Science and Engineering at Flinders University. She is also an adjunct lecturer at the University of Adelaide, an associate investigator with ARC Centre of Excellence for Mathematical and Statistical Frontiers, and a visiting scientist at CSIRO, Australia. She is a member of the Australian Mathematics Society and Women in Maths Special Interest Group.
She did her Ph.D. in Computer Science from University of Konstanz, Germany. Her research lies at the intersection of applied mathematics and social psychology. She is particularly interested in network science, understanding grey-zone tactics, combating online misinformation, serious games, and decision making in the context of cyber security.
She is working on AI-enabled situational understanding models for combating misinformation, using graph-theoretic knowledge-based constructs, coupled with natural language processing techniques and social psychology. Her work has led to the design of agent-based network simulation models that can be deployed in modern wargames which can be used by defence and the government for training decision-makers to combat online misinformation during crisis.
\end{IEEEbiography}

\begin{IEEEbiography}[{\includegraphics[width=1in,height=1.25in,clip,keepaspectratio]{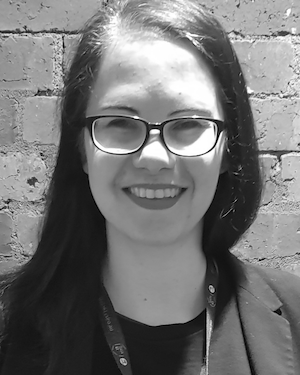}}]{MARTHIE GROBLER} is passionate about making cyber security more accessible. Her research focus is on human centric cyber security, enhancing usability of security solutions by considering human factors. She spearheaded the establishment of a new human centric security research team, which is focused on addressing the alignment and integration of human factors in the cyber domain to enhance security adoption and efficiency. Her expertise falls within a very niche area of cyber security, the intersection between cyber security, usable security and human computer interaction. Marthie has a strong focus on improving cyber security across user groups, considering traditional usability metrics, governance and policies, as well as cyber security maturity and resilience. Her main developments are in the domains of cyber risk and governance, and cyber education and digital upskilling. Marthie currently holds a position as Team Leader: Human Centric Security at CSIRO's Data61 in Melbourne, Australia.
\end{IEEEbiography}

\begin{IEEEbiography}[{\includegraphics[width=1in,height=1.25in,clip,keepaspectratio]{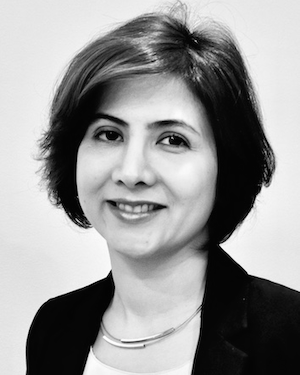}}]{MANSOOREH ZAHEDI} is a lecturer in Software Engineering (SE) at the School of Computing and Information Systems, the University of Melbourne. She has received her PhD from School of Software and Systems at IT University of Copenhagen, Denmark.  
Her research is stimulated by the challenges involved in continuously evolving Software Engineering processes and practices to enable organizations developing high-quality software intensive systems. Her primary research goal is to apply empirical research methods and tools to investigate the role of people, processes, and technologies in different software development paradigms. Her key research interests are human aspects in software engineering, socio-technical aspects of cyber security and continuous software engineering. She has conducted extensive field studies with different companies internationally and published empirically grounded findings. Her work has been published in several high-ranking software engineering venues e.g., EMSE, FSE, JSS, IST, ESEM, EASE. She has served the research community extensively in different capacities, e.g., workshop chair of Evaluation and Assessment in Software Engineering (EASE 2021), Short-papers chair of (EASE 2020), workshop co-chair of international conference on Agile Software Development (XP 2020), poster co-chair of international conference on Model Driven Languages and Systems (MODELS 18, 19), judge at SRC competition (ICSE 2020) and Social activities co-chair of Requirements Engineering conference (RE 2022).
\end{IEEEbiography}

\begin{IEEEbiography}[{\includegraphics[width=1in,height=1.25in,clip,keepaspectratio]{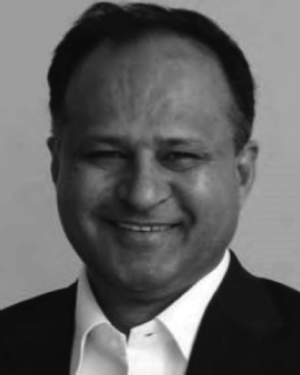}}]{M. ALI BABAR} is currently a Professor with School of Computer Science, The University of Adelaide. He is an Honorary Visiting Professor with the Software Institute, Nanjing University, China. He is also the Director of Cyber Security Adelaide (CSA), which incorporates a node of recently approved the Cyber Security Cooperative Research Centre (CSCRC), whose estimated budget is around AU$\$$140 Millions over seven years with AU$\$$50 Millions provided by the Australia Government. In the area of software engineering education, he led the University’s effort to redevelop a Bachelor of Engineering (software) degree that has been accredited by the Australian Computer Society and the Engineers Australia (ACS/EA). Prior to joining The University of Adelaide, he spent almost seven years in Europe (Ireland, Denmark, and U.K.) as a Senior Researcher and an Academic. Before returning to Australia, he was a Reader of software engineering with Lancaster University. He has established an Interdisciplinary Research Centre, Centre for Research on Engineering Software Technologies (CREST), where he leads the research and research training of more than 30 (20 Ph.D. students) members. Apart from his work having industrial relevance as evidenced by several research and development projects and setting up a number of collaborations in Australia and Europe with industry and government agencies, his publications have been highly cited within the discipline of software engineering as evidenced by his H-index is 52 with 11045 citations as per Google Scholar on December 16, 2021. He leads the theme on Platform and Architecture for Cyber Security as a Service with the CSCRC. He has authored/coauthored more than 220 peer-reviewed publications through premier software technology journals and conferences.
\end{IEEEbiography}

\EOD

\end{document}